\documentclass{ws-ijmpd}

\begin{document}

\markboth{V. Bosch-Ramon \& D. Khangulyan}
{Understanding the Very High-Energy Emission from Microquasars}

%
\catchline{}{}{}{}{}
%

\title{Understanding the Very High-Energy Emission from Microquasars}

\author{Valent\'i Bosch-Ramon}


\author{Dmitry Khangulyan}

\address{Max Planck Institut f\"ur Kernphysik, postfach 10 39 80
69029 Heidelberg, Germany\\
Valenti.Bosch-Ramon@mpi-hd.mpg.de/Dmitry.Khangulyan@mpi-hd.mpg.de}

\maketitle

\begin{history}
\received{Day Month Year}
\revised{Day Month Year}
\comby{Managing Editor}
\end{history}

\begin{abstract}  
Microquasars are X-ray binaries with relativistic jets. These jets are powerful energy carriers,  thought
to be fed by accretion, which produce non-thermal emission at different energy bands. The  processes behind the bulk of the
non-thermal emission in microquasars may be of leptonic (synchrotron and inverse Compton) and hadronic (proton-proton
interactions, photo-meson production, and photo-disintegration) nature. When leptonic, the fast particle cooling would allow
the obtention of relevant information about the properties close to the accelerator, like the radiation and the magnetic
field energy densities, and the acceleration efficiency. When hadronic, the extreme conditions required in the emitter would
have strong implications on the physics of jets and their surroundings. The very high-energy part of the spectrum, i.e.
$>100$~GeV, is a good energy range to explore the physics behind the non-thermal radiation in these compact variable sources.
In addition, this energy range, when taken altogether with lower energy bands, is a key piece to construct a comprehensive
picture of the processes occurring in the emitter. Until recently, the very high-energy range was hard to probe due to the
lack of sensitivity and spatial and spectral resolution of previous instrumentation. Nowadays, however, powerful gamma-ray
instruments are operating and the quality of their observations is allowing, for the first time, to start to understand  the
production of high-energy emission in microquasars.

To date, several Galactic sources showing extended radio emission, among them at least one confirmed microquasar, Cygnus~X-1,
have shown a TeV signal. All of them show complex patterns of spectral and temporal behavior. In this work, we
discuss the physics behind the very high-energy emission in the microquasar Cygnus~X-1, and also in the other two TeV
binaries with detected extended outflows, LS~5039 and LS~I~+61~303, pointing out relevant aspects of the complex phenomena
occurring in them. We conclude that the TeV emission is likely of leptonic origin, although hadrons cannot be discarded. In
addition, efficient electromagnetic cascades can hardly develop since even relatively low magnetic fields suppress them.
Also, the modeling of the radiation from some of the detected sources points to them as either extremely efficient
accelerators, and/or having the TeV emitter at a distance from the compact object of about $\sim 10^{12}$~cm. Finally, we
point out that the role of a massive and hot stellar companion, due to its strong photon field and wind, cannot be neglected
when trying to understand the behavior of microquasars at high and very high energies. The complexity of microquasars
precludes straightforward generalizations to a whole population, and are better studied presently in a source by source base.
The new and future gamma-ray instrumentation 
will imply a big step further in our understanding of the processes in microquasars and gamma-ray emitting binaries.
\end{abstract} 

\section{Introduction}	

Microquasars are an X-ray binary (XRB) subclass formed by those sources that present extended radio jets (e.g. Mirabel \&
Rodr\'iguez \cite{mirabel99}). These systems are formed by a non-degenerated star, which can be in different stages of its
evolution, and a compact object, which can be a black-hole or a neutron star. Depending on the mass of the non-degenerated
stellar companion, the system is considered a low or a high-mass microquasar. Typically, systems harboring an OB star are
considered high-mass XRBs (HMXB), and XRBs with later type stellar companions are classified as intermediate or low-mass
XRBs. It is thought that the compact object powers the relativistic jets via accretion of matter expelled from the companion.
This material, when reaching the surroundings of the compact object, forms an accretion disk that is usually detected in the
X-rays. Simplifying very much the case, depending on the accretion state, the X-ray spectrum varies strongly, from a
multi-color black body peaking around 1~keV plus a minor soft power-law spectrum at higher energies, for high accretion rates
(high-soft state), to one dominated by a hard power-law spectrum plus an exponential cutoff around 100~keV, for low accretion rates (low-hard
state; for an extensive
description of the X-ray phenomenology, see McClintock \& Remillard \cite{mcclintock06}). It is expected that a persistent jet will be present during the low-hard state, and a transient ejection will form
when switching from the low-hard to the high-soft state (Fender et al. \cite{fender04a}). Correlations between the radio and
the X-ray luminosity, and the accretion/ejection activity, have been proposed (e.g. Gallo et al. \cite{gallo03}, Corbel et
al. \cite{corbel03}, Fender et al. \cite{fender03}). 

The jet formation and the production of non-thermal radiation in the jet are major ingredients that distinguish a microquasar
among other types of XRBs. The non-thermal radiation produced in microquasar jets has been resolved in radio at very
different spatial scales (e.g. Rib\'o \cite{ribo05}) and also in X-rays at large scales (e.g. Corbel et al. \cite{corbel02}).
This emission is a clear evidence that particle acceleration takes place in different locations of microquasar jets under
very different conditions (for a discussion on this, see, e.g., Bosch-Ramon \cite{bosch07}). There are also radio emitting
XRBs in which extended emission has not been detected. It has been proposed that these sources could be microquasars as well
(e.g. Fender \cite{fender04}), like the low-mass system XTE~J1118$+$480 (e.g., Chaty et al. \cite{chaty03}), the jet of which
has not been resolved yet. 

Microquasars show up themselves as compact and rapidly variable sources from radio to very high energies. In such a type of
emitters, when radiating particles are leptons, the highest energy part of the spectrum is a good range to explore
non-thermal processes. It is due to the short timescales associated to the particles that produce the emission, which implies
that the accelerator and the emitter are likely the same or similar regions. In case of hadrons, it is possible to derive
important information concerning the jet hadronic content, at least of its relativistic part, whereas at the same time it is
giving information of the conditions of the emitter, like very dense matter and/or target photon fields. Moreover, photons
generated by very high-energy (VHE) electrons and/or protons give a better insight on the mechanism of acceleration and the
conditions under which it takes place, helping to understand better the processes that accelerate particles up to such high
energies. Finally, the presence of a hot and massive star renders a scenario in which photon-photon absorption, and the 
occurrence of electromagnetic cascades, can be studied. This can give important information on the conditions of the massive
star surroundings. 

Historically, the poor spatial resolution and sensitivity of the available instruments working at gamma-ray energies were not
enough for accurate theoretical modeling, although these sources were proposed to be gamma-ray emitters more than  a
decade ago (e.g. Levinson \& Blandford \cite{levinson96}; Levinson \& Mattox \cite{levinson96b}; Paredes et al.
\cite{paredes00}). Nowadays, however, powerful gamma-ray instruments are operating or will start soon, and the quality of
their observations is allowing us, for the first time, to probe the physical processes that take place in microquasars and
their jets. These observations at very high energies give the necessary input to constrain the theoretical models that lacked
in the past. 

The recent evidence of detection of a transient event, at the (post-trial) significance level of 4.1~$\sigma$, from Cygnus
X-1 (Albert et al. \cite{albert07}) have shown that microquasars can indeed produce VHE emission, and VHE observations can
bring to us important information on these sources. Another two interesting cases are LS~5039 and LS~I~+61~303. The former
has been detected in the TeV range by the Cherenkov telescope HESS (Aharonian et al. \cite{aharonian05}), showing a periodic
behavior in the VHE radiation (Aharonian et al. \cite{aharonian06}) with the same period as the orbit (Casares et al.
\cite{casares05}). LS~I~+61~303 has been also detected at TeV energies by the Cherenkov telescope MAGIC, being its emission
variable (Albert et al. \cite{albert06}). Unlike Cygnus~X-1, which is a firmly established microquasar, LS~5039 and
LS~I~+61~303 are no considered at present microquasars due to some peculiarities in their X-ray and radio characteristics.
Nevertheless, both sources present, like Cygnus~X-1, extended radio emission, and harbor OB stars.

In this review, we want to take profit of the new microquasar phenomenology at very high energies, to draw a theoretical
picture of the non-thermal processes that could take place in these sources. This is to be put in the context of the
historical evolution of the field, which will be summarized. To explore which are the relevant processes in the microquasar
scenario, we will carry out a detailed review of different mechanisms: particle transport, radiation and photon-photon
absorption, in the context of microquasars. Using this sound and basic theoretical background, plus the observational
knowledge at very high energies from the sources presented above, the microquasar Cygnus~X-1, and also the TeV binaries
LS~5039 and LS~I~+61~303, constraints on their physical conditions will be inferred. We note that the type of approach
applied to these sources is applicable to large extent to any close binary system emitting in the TeV regime. It is
worth mentioning here a recent review of a broader topic by Levinson (\cite{levinson06}) on VHE radiation from jets including
active galactic nuclei, gamma-ray bursts, and microquasars. It is also interesting to note that
massive young stellar objects, also presenting jets,
have been proposed to be gamma-ray emitters by Araudo et al. (\cite{araudo07}).

This work is organized in the following manner. In Sections~\ref{genst} and \ref{difsc}, we try to look comprehensively at
the microquasar phenomenological picture, setting up a general scenario for these sources. In addition, there we summarize in
a non-exhaustive way previous studies of different topics related to microquasars at high energies. In Sect.~\ref{theor}, a
short review on different particle transport, acceleration and radiation mechanisms is done, treating also briefly the issue
of gamma-ray absorption and electromagnetic cascading in microquasars. All this can help to understand how the new VHE data
fits in previous ideas and frameworks. Also, we put forward a plain but physically sound leptonic model to explore the
processes that are involved in the generation and absorption of VHE emission in jet galactic sources, trying to understand
which kind of physics is relevant there. For this, the recent observational findings at very high energies concerning
microquasars are used. In Sect.~\ref{disc}, some hot topics of the field are discussed with some extension. Finally, in
Sect.~\ref{summ}, the main ideas of this work are summarized.

\section{The study of microquasars}\label{genst}

At the discovery of microquasars, their physics was object of speculation due to the morphological similarities of these
sources with their larger scale analogs, the quasars (Mirabel et al. \cite{mirabel92}). The question was to find out to which
extent the processes taking place in extragalactic jet sources could be extended to galactic ones. In addition to morphology,
kinematical similarities became apparent when superluminal moving ejecta were found at radio wavelengths in GRS~1915$+$150
(Mirabel et al. \cite{mirabel94}). Nowadays it is widely spread the opinion that galactic and extragalactic relativistic jet
sources share much more than just morphological and kinematical resemblance. In fact, since both types of source harbor a
compact object surrounded by an accretion disk and a relativistic outflow that is an efficient non-thermal emitter, the
original analogy argument has been extended not only to the mechanisms that are producing such non-thermal radiation but even
beyond. The aim would be to embrace also the physical link between the relativistic particle generation, the accretion
phenomena, the jet formation and to some extent the interaction between the jet and its environment. A well known example
would be the phenomenological scaling laws relating the physics of the radio and the X-ray emission in microquasars with the
accretion rate (e.g. Corbel et al. \cite{corbel03}; Gallo et al. \cite{gallo03}), and its more general version that includes
the mass of the black-hole, being extended to extragalactic objects (e.g. Merloni et al. \cite{merloni03}; Falcke et al.
\cite{falcke04}; K\"ording et al. \cite{koerding06}). As already mentioned, these empirical scaling laws include the
accretion rate and the central object mass as the key parameters to account for (at some stage Doppler boosting is to be
added), as well as phenomenological links, sometimes accretion and jet formation theory motivated, between the radio, the
X-ray, and the jet kinetic luminosities and the accretion rate itself.

It is usual and seems natural to associate the properties of the accretion disk, the jet and the compact object mass. The
latter will influence strongly the properties of the infalling matter, like the accretion rate, the accreted matter
temperature, the magnetic field strength or the accretion disk radiative output. In many XRBs, standard accretion theory can
explain the X-ray spectrum quite accurately via thermal emission from the inner parts of the accretion disk plus a
corona-like emitter\footnote{A region surrounding the compact object that would be filled with a hot population of electrons
and ions.}, although there is an on-going debate concerning the origin of the hard X-rays, whether they come from the jet
base, or from a corona-like region close to the compact object, both of similar properties (e.g. Markoff et al.
\cite{markoff05}; Maccarone \cite{maccarone05}). In any case, some sources do not fit in any of these schemas, since their
X-ray radiation does not seem to come from the regions close to the compact object but further out, like the case of SS~433
and, perhaps as well, LS~5039 and LS~I~+61~303. This could be explained by recent theoretical studies that show that the role
of the magnetic field can be more crucial than the compact object mass itself and can lead to radiatively inefficient
accretion (Bogovalov \& Kelner \cite{bogovalov05}), in which case the jet could dominate in the X-ray band (see also Fender
et al. \cite{fender03}).

Theoretically, the relationship between the compact object mass and the jet physics (e.g. ejected mass rate, matter content,
internal energy, carried magnetic energy, bulk speed) is unclear since the accretion/ejection physics is not well known.
Furthermore, the link between the main jet properties and the production of non-thermal particles and their radiation cannot
be based on first principles, since it relies on the particular conditions of the jet plasma (sound speed, diffusion
coefficients, local magnetic field, degree of turbulence, presence of shocks and their velocity, radiation field, plasma
density and temperature, etc.), which are not properly constrained. Actually, the situation is even more complicated due to
the influence on the jet processes and the non-thermal radiation itself of external factors like, for instance, an external
radiation field, the wind produced by the star, the properties of the interstellar medium (ISM), etc. Therefore, although
empirical laws can help to classify objects via significant features of their emission and be powerful heuristic tools, it is
sometimes difficult to motivate them from first principles, and fundamental approaches become necessary. Such fundamental
approaches have to be applied source by source and grounded on detailed observational data. It requires keeping a
comprehensive approach to study the source (i.e. phenomenological studies), but developing more fundamental theoretical
models.

\section{High energy processes in microquasars: a historical perspective}\label{difsc}

In this section, a schematic and descriptive picture for the relevant processes taking place in microquasars, like particle acceleration, and generation and absorption of VHE
radiation, is presented together with a historical perspective of the field. This will set up the context within which a more detailed physical treatment will be carried out in
Sect.~\ref{theor}.

\subsection{A descriptive picture}

\begin{figure}[pb]
\centerline{\psfig{file=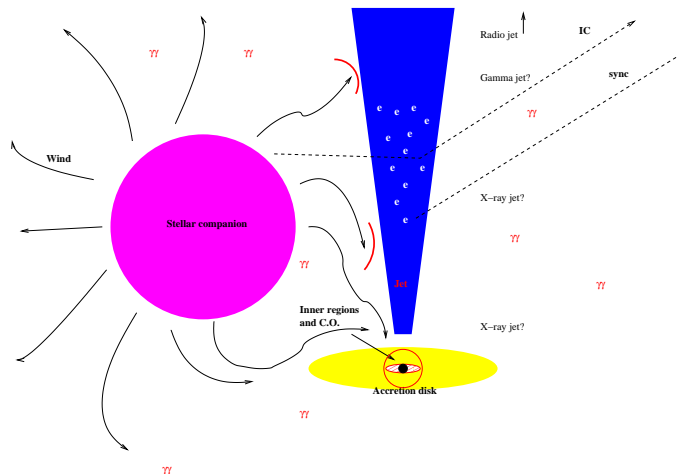,width=9cm}}
\vspace*{8pt}
\caption{A sketch of a microquasar. We explicitly show the star, the stellar wind, the accretion
disk, the jet, the jet relativistic electrons, some leptonic radiation processes, and pair creation
under the stellar photon field.\label{mq}}
\end{figure}

For illustrative purposes, we show in Figure~\ref{mq} a very schematic picture of a microquasar together with its main
elements, and some of the processes taking place in these objects. Some regions of the jet are labeled depending on the main
radiative products. This simplified picture would consist on several important elements: the companion star, the stellar
radiation field, the stellar wind, the compact object, the accretion disk, and the jet itself. The relativistic particles
inside the jet will interact with the stellar radiation field as well as the jet magnetic field producing inverse Compton
(IC) and synchrotron radiation, respectively. Under the strong stellar radiation field, the creation of pairs will be
unavoidable if VHE gamma-rays are produced. Finally, the stellar wind may have a significant impact on the jet. All this,
plus some additional physical processes (not included in Fig.~\ref{mq} for clarity), like hadronic emission, will be
discussed in more detail in Sect.~\ref{theor}. 

\subsection{Jet formation, evolution and termination}

We present here a brief description of a plausible picture for the jet development, from the launching to the termination
point. 

Although a complete theory for jet generation is still lacking, many studies on jet powering, acceleration and collimation
have been carried out during the last decades (e.g. Blandford \cite{blandford76}; Blandford \& Znajek \cite{blandford77};
Blandford \& Payne \cite{blandford82}; Meier \cite{meier96}; Koide et al. \cite{koide02};  Chattopadhyay \& Chakrabarti
\cite{chatto02}; Meier \cite{meier03}; Hujeirat \cite{hujeirat04}; Meier \cite{meier05}; Bogovalov \& Kelner
\cite{bogovalov05}; De Villiers et al. \cite{deVilliers05}; Ferreira et al. \cite{ferreira06}; 
McKinney \cite{mcKinney06};
Hawley \& Krolik \cite{hawley06};
Komissarov et al. \cite{komissarov07}, Barkov \& Komissarov \cite{barkov08}). At present, due to the apparent correlation between accretion and jet activity (e.g.
Fender et al. \cite{fender04a}), a widely accepted scenario is one in which jets are powered by accretion. The accreted
matter starts to move following ordered magnetic field lines that thread the inner regions of the accretion disk. By
magneto-centrifugal forces, the plasma is ejected by these magnetic field lines from the accretion disk in the direction
perpendicular to it. The differential rotation of the accretion disk would create a spiral-like shape of the magnetic lines.
This magnetic field  configuration would accelerate and collimate the plasma. Although it is unclear at which scales the jet
is already formed, VLBI observations of extragalactic jets show evidences that the collimation region could be located at
$\sim 100$--1000~$R_{\rm Sch}$ (e.g. Junor et al. \cite{junor99}; Horiuchi et al. \cite{horiuchi06})\footnote{The jets of the
microquasar SS~433 present collimation distances, inferred from X-ray observations, apparently larger, of about $\sim
10^6$~$R_{\rm Sch}$ (Namiki et al. \cite{namiki03}). We note however that the jet of SS~433 is quite different from a typical
compact microquasar jet because of its huge kinetic power, messy environment, and thermal emitting nature.}.

Once the microquasar jet is formed, it may interact with dense material ejected by the accretion disk or the stellar
companion, the latter being almost unavoidable in high-mass systems. Although the role of an accretion disk wind could be to
give further collimation and stability to the jet (e.g. Hardee \& Hughes \cite{hardee03}; Tsinganos et al.
\cite{tsinganos04}), the role of a strong lateral stellar wind may lead to jet bending and even disruption (Perucho \&
Bosch-Ramon \cite{perucho07}). In any case, jet/environment interaction can lead to shock formation and the radiative
counterpart may be observable either as a transient phenomena, when the jet (or a discrete ejection) penetrates for the first
time the surrounding medium, or as a (quasi) steady one, when the jet formation is continuous at the relevant timescales and
recollimation shocks occur (Perucho \& Bosch-Ramon \cite{perucho07}).

Despite the environment of microquasars at large spatial scales may be quite different depending on the Galaxy region or the
strength of the companion star wind, the jets should stop somewhere, terminating either via disruption, in a similar way to
extragalactic Fanaroff-Riley I sources (e.g. Fanaroff \& Riley \cite{faranoff74};  Perucho \& Mart\'i \cite{pm07}), or via
strong shocks in the ISM, as seems to be the case for Fanaroff-Riley II sources (e.g Fanaroff \& Riley \cite{faranoff74};
Kaiser \& Alexander \cite{kaiser97}; Scheck et al. \cite{sch02}). In either case, the radiative outcomes would be different.
A classical example of the interaction between a microquasar jet and its environment at large scales is the case of the
W~50-SS~433 system (e.g. Zealey et al. \cite{zealey80}; Vel\'azquez \& Raga \cite{velazquez00}).

\begin{figure}[pb]
\centerline{\psfig{file=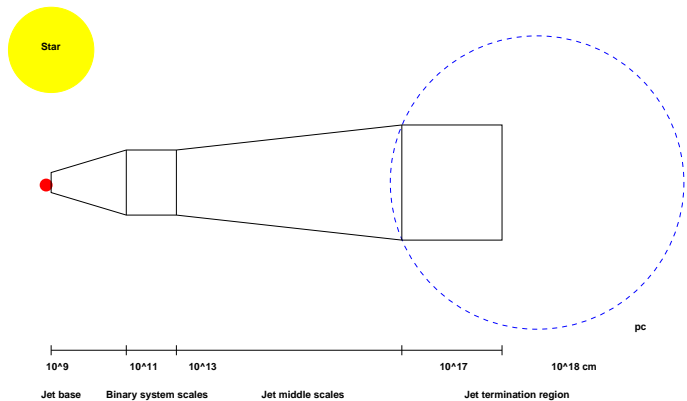,width=9cm}}
\vspace*{8pt}
\caption{Sketch of a microquasar with the different relevant scales resumed in this section. \label{mqq}}
\end{figure}

In the following sections, we list several non-thermal processes, i.e. particle acceleration and radiation, that may occur in
the jet giving a (non exhaustive) overview on the literature. For the sake of clarity, we have divided the jet in several
regions. The jet base, close to the compact object ($\sim 100-1000$~R$_{\rm Sch}$), a region farther out, at the binary
system scales (typically $\sim 10^{11}-10^{13}$~cm), well outside the binary system -jet middle scales- (around
$10^{15}-10^{16}$~cm), and the termination regions of the jet, where it ends interacting somehow with the ISM ($\ge
10^{17}$~cm). We show in Fig.~\ref{mqq} a sketch of the different considered regions. Later on, with the most recent
observational findings at hand, we will go deeper in the modeling to find out which of the different considered scenarios is
the most suitable to explain the data.

\subsection{Particle acceleration}

Particle acceleration processes in microquasar jets can be of different types. At the base, the jet could be magnetically
dominated (e.g., in extragalactic jets: Sikora et al. \cite{sikora05}), and a mechanism of particle acceleration may be
magnetic energy dissipation via MHD instabilities in the jet base (e.g. Zenitani et al. \cite{zenitani01}). Also, if jet
velocities were high enough in the base, the dense available photon (from an accretion disk/corona) and matter fields could
allow the converter mechanism to take place (e.g. Derishev et al. \cite{derishev03}; Stern \& Poutanen \cite{stern06}).
Magnetic field reconnection in the surrounding corona could inject a non-thermal population of particles in the jet as well
(e.g. Gierlinski \& Done \cite{gierlinski03} and references therein). A magneto-centrifugal mechanism could also operate very
close to a rotating black-hole (e.g. Neronov \& Aharonian \cite{neronov07}; Rieger \& Aharonian \cite{rieger07}).

At binary system scales, plausible mechanisms to generate relativistic particles in the jet are the different versions of the
Fermi process: shock diffusive (Fermi~I); random scattering (Fermi~II); and shear acceleration (e.g. Drury \cite{drury83};
Fermi \cite{fermi49}; and Rieger \& Duffy \cite{rieger04}, respectively; see also Rieger et al. \cite{rieger06}). Fermi~I
mechanism could take place due to internal shocks in the jet; Fermi~II acceleration could take place if magnetic turbulence
is strong enough, with high Alfven velocity; shear layer would be a natural outcome of an expanding jet or different
jet/medium velocities. Interactions with the stellar wind may also trigger particle acceleration via, e.g., a recollimation
shock formed in the jet that expands against the dense material expelled by the companion star (e.g. Perucho \& Bosch-Ramon
\cite{perucho07}). The velocities of the shocks mentioned here could be either mildly or strongly relativistic. In the latter
case, the converter mechanism may be effective in very bright star systems.

At microquasar jet middle scales, some sort of shock acceleration might still take place. For instance, 
intermittent ejections 
at timescales of $\sim$hours--days and different velocities could create shocks at distances of about $\sim 10^{15}-10^{16}$~cm. Also
Fermi~II type and shear acceleration appear plausible for a continuous outflow at these scales (something similar could
happen in the intra-knot regions of extragalactic jets, see, e.g., Rieger et al. \cite{rieger06}). Regarding the environment
dynamical influence on the jet, it is not expected to be significant given the high jet ram pressure compared with the medium one
at these scales.

At the microquasar jet termination point, as in AGN hot spots and radio lobes (e.g. Kaiser \& Alexander \cite{kaiser97}), the
external medium plays an important dynamical role. When the swept ISM inertia starts to affect the jet advance, two shocks may
be formed, one moving backwards in the jet, the so-called reverse shock, and another one moving forward, the so-called
forward or bow shock. Under these conditions, Fermi~I type acceleration mechanism seems the most reasonable option, although
high diffusive and convective rates in the downstream regions of the forward/reverse shocks could prevent efficient
acceleration to occur. It might be the case as well that hydrodynamical instabilities distorted the jet and mixed jet matter
with the ISM without forming strong shocks (e.g. Heinz \& Sunyaev \cite{heinz02}).

\subsection{Radiative processes}

In the jet base, depending on the dominant conditions, the relevant leptonic radiative mechanisms could be synchrotron
emission (e.g. Markoff et al. \cite{markoff01}), relativistic Bremsstrahlung from electrons interacting with jet ions
(Bosch-Ramon et al. \cite{bosch06b}), SSC (e.g. Bosch-Ramon \& Paredes \cite{bosch04}) and IC with corona and/or disk photons
(e.g. Romero et al. \cite{romero02}; Georganopoulos et al. \cite{georganopoulos02}). Regarding hadronic processes, there are
several radiative mechanisms that could produce gamma-rays, neutrinos and, as a by-product, low energy emission from
secondary pairs. Two of these mechanisms are the collisions of relativistic protons with ions (pp) in the jet, and
interactions between jet relativistic protons and X-ray photons (photo-meson) from the disk, the corona or the jet itself
(e.g. Levinson et al. \cite{levinson01}; Aharonian et al. \cite{aharonian06b}; Romero \& Vila \cite{romero08} -who also
account for proton synchrotron emission-). These relativistic proton collisions with ions and photons would produce neutral
pions ($\pi^0$) that decay to gamma-rays, and charged pions ($\pi^\pm$) that decay to muons and neutrinos, the former
decaying then to electron-positron pairs and neutrinos. Another possible hadronic mechanism is photo-disintegration, which
requires the presence of UHE heavy nuclei and a dense field of target photons of large enough energy. This process produces
lower mass hadrons and gamma-rays.

At binary system scales, possible radiative leptonic processes taking place in microquasars are synchrotron emission (e.g.
Yuan et al. \cite{yuan05}; Paredes et al. \cite{paredes06}), relativistic Bremsstrahlung (e.g. Bosch-Ramon et al.
\cite{bosch06b}), SSC (e.g. Atoyan \& Aharonian \cite{atoyan99};  Dermer et al. \cite{dermer06}), and external IC (e.g.
Paredes et al. \cite{paredes00,paredes02}; Kaufman Bernad\'o et al. \cite{kaufman02}; Georganopoulos et al.
\cite{georganopoulos02}; Dermer et al. \cite{dermer06}; Khangulyan et al. \cite{khangulyan07}). At these spatial scales, jet
proton collisions with target nuclei of the stellar wind (e.g. Romero et al. \cite{romero03}; Aharonian et al.
\cite{aharonian06b}) seem to be the most efficient hadronic process. As noted above, this mechanism would lead as well to
neutrino production (e.g. Romero et al. \cite{romero05}; Aharonian et al. \cite{aharonian06b}). Other hadronic processes,
which were also discussed in the literature, are photo-meson production (e.g. Aharonian \cite{aharonian06c}) and
photo-disintegration (e.g. Bednarek \cite{bednarek05}).

The emission at larger scales are commonly characterized by synchrotron radiation. At higher energies, stellar IC scattering
is quite inefficient because the large distances to the companion star and the subsequent dilution of the stellar photon
field. Nevertheless, for powerful ejections, SSC could still be significant (e.g. Atoyan \& Aharonian \cite{atoyan99}).
Regarding the particle energy distribution, its evolution is likely dominated by convective and adiabatic energy losses (van
der Laan \cite{vanderlaan66}). At the termination of the jet, in case particle acceleration and confinement were efficient,
synchrotron, relativistic Bremsstrahlung and IC radiation could be produced and even detected for galactic sources (e.g.
Aharonian \& Atoyan \cite{aharonian98}; Bordas et al. \cite{bordas08}). Hadronic acceleration could take place as well, which
could lead to gamma-ray production (e.g. Heinz \& Sunyaev \cite{heinz02}) and secondary leptonic emission (e.g. Bosch-Ramon
et al. \cite{bosch05}).

Regarding the variability of the emission discussed above, several factors are relevant: injection could change via
variations in the accelerator (e.g. injection power, injection spectrum); the target densities could vary via changes in the
magnetic, photon and matter fields; geometry changes, due to e.g. orbital motion or jet precession, could affect anisotropic
gamma-gamma absorption and IC scattering or the level of radiation Doppler boosting. Thus, the timescales of the variability
could be linked to the injection mechanism, the radiative cooling, particle escape, and the orbital motion, depending on
which mechanism plays a major role. Depending on the emitter location and size, some kinds of variability will not play a
role because they will be smeared out. We come back to this issue for a more detailed discussion in Sect.~\ref{theor}. 

\subsection{Photon-photon absorption and electromagnetic cascades}

Close to the compact object, either in the inner accretion disk, in the corona-like region, or in the jet base, quite extreme
conditions should be present. Magnetic fields are likely high, and the same applies to the present photon fields, which may
be dominated by thermal radiation from accretion peaking at UV/X-rays. Dense UV/X-ray photon fields imply that $\sim$~GeV
gamma-ray absorption close to the compact object will be very high. The strong magnetic field may suppress efficient
electromagnetic cascading, although the occurrence of cascades (e.g. Akharonian et al. \cite{akharonian85}) cannot be
discarded. At binary system scales, the photon-photon opacities of the stellar photon field for gamma-rays are high in
massive systems. Although high-mass stars can have quite large magnetic fields in their surfaces, up to $10^3$~G for a young
O star (e.g. Donati et al. \cite{donati02}), electromagnetic cascades may still develop in the system since the magnetic
field strength at several stellar radii is not well known. Several authors have studied the photon-photon absorption effects
in the VHE lightcurve (e.g. Protheroe \& Stanev \cite{protheroe87}; Moskalenko \& Karakula \cite{moskalenko94}; Bednarek
\cite{bednarek97}; Boettcher \& Dermer \cite{boettcher05}; Dubus \cite{dubus06}; Khangulyan et~al. \cite{khangulyan07};
Reynoso et~al. \cite{reynoso08}). The effects of absorption on the radiation variability are important not only because the
photon column density changes along the orbit, but also due to the angular dependence of the cross section and the low-energy
threshold of the pair creation process. The interaction angle between gamma-rays and stellar photons changes with the orbital
phase. Several studies have been done in the recent years studying the impact of cascading in the gamma-ray spectrum of
microquasars. Some authors have assumed that particles get deflected after creation, using a three-dimensional code to
compute cascading, (e.g. Bednarek \cite{bednarek06}), whereas others have computed one-dimensional cascades in the direction
to the observer (e.g. Aharonian et al. \cite{aharonian06b}; Orellana et al. \cite{orellana07}; Khangulyan et al.
\cite{khangulyan07}). In case the magnetic field is high enough, synchrotron secondary radiation produced in the system will
be significant in the radio and X-ray domains (e.g. Khangulyan et al. \cite{khangulyan07}; Bosch-Ramon et al.
\cite{bosch08}). In the next section, we discuss further this issue.

\section{Theoretical interpretation of observations}\label{theor}

To interpret observations, one can use some analogies comparing one specific object with the general properties of a
population of sources. Unfortunately, neither our knowledge on the physics of microquasars emitting at TeV, nor the number of
these sources, are sufficient to follow this strategy. Another approach, which seems to us more suitable to study the
complexity of the behavior of these objects at very high energies, is to start from basic physical processes to understand
the observations in a rough but sound way. It implies to check, with the help of data and the available knowledge on each
particular source, which are the most reasonable physical conditions, and processes, that lead to the observed VHE emission.
To perform such an analysis, it is required to constrain the conditions in which particle acceleration is possible up to the
observed energies, to know which mechanism is behind gamma-ray production (either leptonic -IC- or hadronic -pp, p$\gamma$
and nuclei disintegration-), and the impact of photon-photon absorption (and the subsequent energy release channel
-synchrotron or IC-). In addition, it is important to explore the impact of particle transport (advection, diffusion or
particle escape) on the final spectrum. In the following, each one of these elements is considered individually. It is worth
to remark that in reality the mentioned processes will be likely coupled, and a very complex outcome can be expected, as
shown in Sects.~\ref{mQc}, \ref{ls} and \ref{lsi}, when focusing in the leptonic scenario.


\subsection{Non-thermal processes}\label{nonther}

In this section, we explore those processes that are relevant to understand the non-thermal emission produced at very high
energies for the expected conditions given in microquasars. We do not focus particularly in the jet but also in its
environment. In particular, we consider particle acceleration, particle energy losses and transport, and photon-photon
absorption and electromagnetic cascading.

\subsubsection{Particle acceleration}

Our aim is not to model accurately what is observed, but to derive basic and solid constraints from the available data. The
first step is to set a necessary condition for acceleration of a charged particle of a certain energy to occur, which is
given by the Hillas criterium (Hillas \cite{hillas84}). This consists on the fact that particles can only be accelerated if
their Larmor radius  ($r_{\rm L}=E/qB_{\rm a}$; where $B_{\rm a}$ is the accelerator magnetic field, and $q$ and $E$ are the
charge and energy of the particle, respectively) is smaller than the accelerator size ($l_{\rm a}$), since otherwise they
will escape the accelerator. This limits the highest achievable energy to: 
\begin{equation}
E<qB_{\rm a}l_{\rm a}\,,
\label{larm}
\end{equation}
which can be rewritten in the following form:
\begin{equation}
E<30\,\frac{q}{e}\,B_{\rm a,G}\,l_{11}\, {\rm TeV}\,,
\label{maxelarm}
\end{equation}
where $e$ is the electron charge, $B_{\rm a,G}$ is the accelerator magnetic field ($B_{\rm a}$) in Gauss units, 
and $l_{11}=l_{\rm a}/10^{11}{\rm cm}$.
Nevertheless, to determine whether particles can be accelerated up to a certain energy, the specific acceleration and 
energy loss (or particle escape)
mechanisms are to be known: $t_{\rm acc}=t_{\rm cool/esc}$\,. In general, the acceleration timescale can be expressed as:
\begin{equation}
t_{\rm acc}=\eta{r_{\rm L} \over c}\simeq 0.1\eta \frac{e}{q}E_{\rm TeV}B_{\rm a,G}^{-1}\,{\rm s}\,,
\label{accrate}
\end{equation}
where $\eta$ is a dimensionless phenomenological parameter (or function) representing the acceleration efficiency, and $E_{\rm
TeV}$ is the particle energy in TeV units. The particular case of $\eta=1$ corresponds to the shortest possible acceleration time
independently of the acceleration mechanism. Another instance for $\eta$ can be given for the case of
non-relativistic diffusive shock acceleration (plane shock with weak magnetic field, in the test particle
approximation) (e.g. Protheroe \cite{protheroe99}):
\begin{equation}
\eta=2\pi{D\over D_{\rm Bohm}}\left(c\over v_{\rm sh}\right)^2\,,
\label{accef}
\end{equation}
where $ v_{\rm sh}$ is the shock velocity, and $D$ is the  diffusion coefficient ($D_{\rm Bohm}$ in the Bohm limit). For $v_{\rm
sh}=3\times 10^9$~cm~s$^{-1}$ and $D=D_{\rm Bohm}$, $\eta\sim 10^3$.

\begin{figure}[pb]
\centerline{\psfig{file=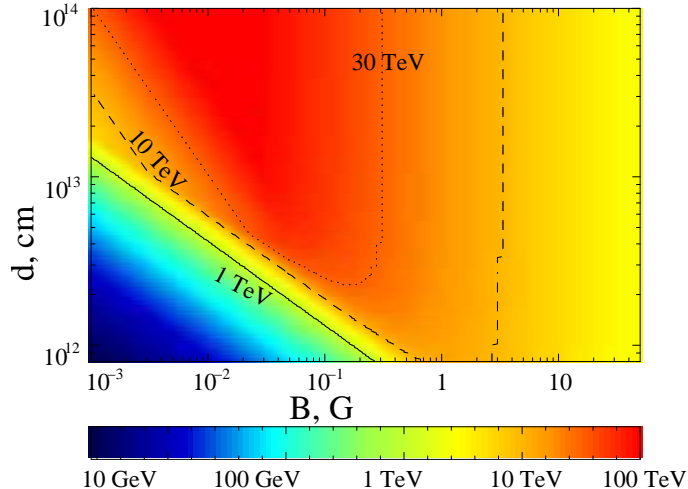,width=9cm, angle=0}}
\vspace*{8pt}
\caption{
2-dimensional $d-B$ map that shows the maximum achievable energy for $\eta=10$ for
different $d_{\rm a}$ and $B_{\rm a}$ values. The adopted parameter values are $L_*=10^{39}$~erg~s$^{-1}$, 
$d_*=3\times 10^{12}$~cm and $kT_*\approx 3$~eV. \label{acc}}
\end{figure}

As an example of the importance of using acceleration constraints, we show in Fig.~\ref{acc} a 2-dimensional  $d_{\rm
a}-B_{\rm a}$ map for a compact high-mass microquasar with stellar luminosity $L_*=10^{39}$~erg~s$^{-1}$, $d_*=3\times
10^{12}$~cm and $kT_*\approx 3$~eV. $d_{\rm a}$ is the distance from the accelerator to the star. As done by Khangulyan et
al. \cite{khangulyan07}, it is possible to restrict $d_{\rm a}$ and $\eta$. In the case of LS~5039, for instance, for
reasonable $\eta\ge 10$, and the detected 30~TeV photon energies, the accelerator should be outside the system, i.e. 
$d_{\rm a}> 2\times 10^{12}$~cm.

{\bf Particle transport}

Without focusing on any particular acceleration mechanism, we note that, either the maximum energy is limited by cooling, 
which implies that particles 
radiate most of their energy before escaping the accelerator, or their escape stops acceleration. 
In the
latter case, the emitter itself can be considered 
larger than the acceleration region. The escape time ($t_{\rm esc}$) can be characterized, in
the accelerator as well as in the whole emitter, by the minimum among 
the diffusion ($t_{\rm diff}$) and the advection ($t_{\rm adv}$) times:
\begin{equation}
 t_{\rm esc}=\min(t_{\rm diff}, t_{\rm adv})\,,
\end{equation}
which can be expressed either as:
\begin{equation}
 t_{\rm diff}={l^2\over2D(E)}\sim 2\times 10^4 l_{\rm 12}^2
 \left(\frac{D_{\rm Bohm}}{D(E)}\right)B_{\rm G} E_{\rm TeV}^{-1}\,{\rm s}\,,
\end{equation}
for 1-dimensional diffusion in the accelerator/emitter, 
where $l$ ($l_{12}=l/10^{12}\, {\rm cm}$) is the (accelerator/emitter) size covered by diffusion,
and $B_{\rm G}$ is the emitter magnetic field; or as:
\begin{equation}
 t_{\rm adv}={l\over V_{\rm adv}}\sim 10^4 l_{\rm 12}V_{8}^{-1}\,{\rm s}\,,
\end{equation}
for advection, where $V_{\rm adv}$ is the advection speed ($V_{8}=V_{\rm adv}/10^{8}$~cm~s$^{-1}$).

Under the impact of cooling, the typical distance up to which particles can propagate is:
\begin{equation} 
l_{\rm diff}=10^{10}\,E_{\rm TeV}^{1/2} B_{\rm G}^{-1/2}t_{\rm cool}^{1/2}
\left(\frac{D}{D_{\rm Bohm}}\right)^{1/2}\,\,{\rm cm}\,,
\label{diffs}
\end{equation}
under diffusive transport, and:
\begin{equation} 
l_{\rm adv}=10^{10}\left(\frac{V_{\rm adv}}{10^{10}\,{\rm cm/s}}\right)t_{\rm cool}\,\,{\rm cm}\,,
\label{advs}
\end{equation}
under advective transport. $t_{\rm cool}$ is the cooling timescale (in $s$) of the dominant loss mechanism.
We discuss in the following different cooling processes that may be
relevant in the microquasar context.

\subsubsection{Radiative processes}\label{rad}

{\bf Synchrotron}

In the presence of a disordered magnetic field, electrons radiate via synchrotron emission. 
The synchrotron cooling timescale 
is approximately:
\begin{equation} 
t_{\rm sy}\approx 4\times 10^2 \left(\frac{e}{q}\right)^4(m/m_{\rm e})^4B_{\rm a,G}^{-2} E_{\rm TeV}^{-1}\, {\rm s}\,, 
\label{sy_loss}
\end{equation}
where $m$ and $m_{\rm e}$ are the mass of the particle and the electron, respectively. 
This gives the maximum particle energy:
\begin{equation} 
E_{\rm max}  \approx 60 \, \left(\frac{e}{q}\right)^{3/2}(m/m_{\rm e})^2\,(\eta B_{\rm a,G})^{-1/2} \, {\rm TeV}\,.
\label{syn_max_en}
\end{equation}
This process is only relevant for leptons unless the magnetic field is very strong, 
in which case hadronic synchrotron may become efficient.

To see what may happen in a microquasar regarding $E_{\rm max}$ under synchrotron cooling, we take $B_{\rm 0,G}=10^6$~G as a reasonable value for the jet base  magnetic field (at
$\sim 50\,R_{\rm Sch}\sim 10^8$~cm from the compact object), obtaining: $E_{\rm max}\approx 0.06\,\eta^{-1/2}$~TeV. If we assume a distance dependence of $B$ of the form $1/Z$ and
locate the accelerator at $10^{12}$~cm, we obtain: $E_{\rm max}\approx 6\,\eta^{-1/2}$~TeV, i.e. in such a situation, it seems unlikely to produce VHE leptons in the jet even at
the system scales.

At spatial scales similar or smaller than the binary system, it seems reasonable to expect large magnetic fields related either to the jet, to the accretion disk, or to the
companion star. It implies that synchrotron emission can be an efficient radiation process, and electrons may release most of their energy via synchrotron emission. At larger
scales, if no significant magnetic field enhancement takes place, efficiencies will decrease strongly.

It is worthy using Eqs.~(\ref{diffs}) and (\ref{advs}) to calculate the impact of synchrotron losses on the propagation of
electrons via diffusion and advection, since they tell how far relativistic electrons can go without assuming additional acceleration and
neglecting other sources of losses. Under the next parameter choice, e.g. $B\sim 1$~G,  $V_{\rm
adv}=10^{10}$~cm~s$^{-1}$, $D=D_{\rm Bohm}$, advection is the most effective transport mechanism, and TeV particles 
may reach distances of $\sim 3\times 10^{12}$~cm from their injection point. 

{\bf Inverse Compton} 

Under the radiation field of the primary star in a high-mass microquasar, or in the case of strong accretion disk/corona
emission, IC scattering is to be considered and may limit particle acceleration. Synchrotron self-Compton could become a
dominant process, although it is not treated here, since it would require significantly large magnetic fields, making
acceleration up to very high energies unlikely due to strong synchrotron/Thomson IC energy losses. In addition, once the
electron enters in the KN IC regime, the interaction efficiency reduces strongly and synchrotron cooling becomes dominant,
making SSC not very efficient in the TeV range under UV photon fields. We note that, due to the same reason as in the
synchrotron case, IC losses are only significant for leptons\footnote{For hadrons, other cooling processes, mainly
photo-meson production, discussed further in the text, will overcome hadronic IC scattering.}.

For a Planckian distribution of target photons with temperature $T$, the IC energy loss rate can be approximated,
with an accuracy of less than a 3\%, by:
\begin{equation}
\dot{\gamma}_{IC}=5.5\times 10^{17}T_{\rm mcc}^3\gamma {ln(1+0.55\gamma T_{\rm mcc})\over 1+25T_{\rm mcc}\gamma}
\left(1+{1.4\gamma T_{\rm mcc}\over1+12\gamma^2T_{\rm mcc}^2}\right)\,{\rm s}^{-1},
\label{gammabb}
\end{equation}
where $T_{\rm mcc}=kT/m_{\rm e}c^2$.

Regarding the particle maximum achievable energy, at the energies in which we are interested in this work,
IC scattering proceeds in the Klein-Nishina (KN) regime ($\gamma\gg 1/kT$ in $m_{\rm e}c^2$ units). 
In such a case, 
a simple power-law fit for the cooling time can be used for a black-body type of target photon distribution: 
\begin{equation}
t_{\rm IC}\approx 10^2\,
\left(R\over R_*\right)^2
T_{4}^{-2.3}\,
E_{\rm TeV}^{0.7}\,{\rm s}.
\label{tkn1}
\end{equation}
where $R$ and $R_*$ are the distance to the origin and the radius of the source of target photons, 
and $T_4=T_*/10^4\, {\rm K}$.
Eq.~(\ref{tkn1}) can be expressed in a more convenient form in case of a hot and massive primary star as the 
dominant source of target photons:
\begin{equation}
t_{\rm IC}=10^2
(w_{100})^{-1}
\left(T_*\over 3\times 10^4\,{\rm K}\right)^{1.7}\,
E_{\rm TeV}^{0.7}\,{\rm s}\,,
\label{tkn2}
\end{equation}
where $w=100\,w_{100}$~erg~cm$^{-3}$ is the target photon field energy density, and $T_*$ the stellar temperature.
This expression gives a reasonable agreement for the IC cooling time, within a factor $\le 2$, in the relevant energy range: 
\begin{equation}
0.1\ {\rm TeV}
\left(T_{\rm star}\over 3\times 10^4{\rm K}\right)^{-1}<
E<
3\times 10^3\ {\rm TeV}\left(T_{\rm star}\over 3\times 10^4{\rm K}\right)^{-1}\,.
\end{equation}
The corresponding value of the maximum energy is:
\begin{equation}
E_{\rm max}\approx
4\times 10^{10} \, 
[B_{\rm a~G}\eta^{-1}w_{100}^{-1}]^{3.3}~{\rm TeV}\,,
\label{emax}
\end{equation}
which shows that KN IC limits particle acceleration much less than synchrotron radiation, and can be hardly dominant
at the maximum particle energy for reasonable magnetic fields in any region of the microquasar.

To estimate the impact of IC losses on the propagation of electrons via diffusion and advection, Eqs.~(\ref{diffs}) and
(\ref{advs}) can be used as well. Like in the case of synchrotron losses, for the conditions $B\sim 1$~G, $V_{\rm
adv}=10^{10}$~cm~s$^{-1}$ and $D=D_{\rm Bohm}$, and IC loss dominance, advection is again the most efficient transport
mechanism, under which TeV particles could reach distances up to $\sim 10^{12}$~cm.

{\bf Proton-proton interactions}

As mentioned in the previous section, purely hadronic processes like pp interactions
have been discussed in the past in the context of microquasars. We consider them here as well, 
since they may be relevant in some cases. The energy loss timescale for pp collisions (Kelner et al. \cite{kelner06}) is:
\begin{equation}
 t_{\rm pp}\approx {10^6\over n_9}\, {\rm s}\,,
\label{tpp}
\end{equation}
where $n_9=n_{\rm t}/10^9~{\rm cm}^{-3}$ is the target density. 
The energy threshold of this process in the reference frame of the interaction center of masses
is the pion rest mass, $\approx 140$~MeV.
From Eqs.~(\ref{accrate}) and (\ref{tpp}), the maximum particle energy can be derived:
\begin{equation}
E_{\rm max}\approx 10^7 {B_{\rm a,G}\over \eta n_9}\, {\rm TeV}\,.
\label{accrate}
\end{equation}
Given the long cooling timescale, the maximum energy will be likely 
limited by the accelerator size. 

Defining $L_{\rm p}$ as the luminosity injected in the form of relativistic protons, the
luminosities in gamma-rays, neutrinos and secondary electron-positron pairs are,
with differences of about a factor of 2 (Kelner et al. \cite{kelner06}), the next ones:
\begin{equation}
L_{\gamma}\approx \min(1,t_{\rm esc}/t_{\rm pp})\,c_{\rm pp}\,L_{\rm p}\,,
\label{lumpp}
\end{equation}
where $c_{\rm pp}$ is the energy transfer efficiency from relativistic protons to secondary particles ($\sim 10$\%).
In the context of high-mass microquasars, a reasonable lower limit for $t_{\rm esc}$ is the wind advection time, i.e. the
time required for the stellar wind to cross the orbital radius ($R_{\rm orb}/V_{\rm w}$; where $R_{\rm orb}\sim 10^{12}-10^{13}$~cm
and $V_{\rm w}\sim 1-3\times 10^8$~cm~s$^{-1}$), $t_{\rm esc}\le 10^4$~s; and a very lower limit is set by
the speed of light, i.e. $t_{\rm esc}\ge 10^2$~s. All this, plus adopting a $\dot{M}=10^{-6}$~M$_{\odot}$~yr$^{-1}$ 
(typical for O stars), yields:
$$
L_{\gamma}\sim 10^{-5}-10^{-3}L_{\rm p}\,.
$$
Larger efficiencies cannot be discarded in some specific cases, like density enhancements in the wind or even in the
jet itself via, e.g., shocks. Once relativistic protons leave the binary system, the expected density decreases in the
stellar wind as well as in the jet, making pp collisions negligible.

{\bf Photo-meson production}

Among different hadronic processes, photo-meson production (Kelner et al. \cite{kelner08}) is worthy also to be considered.
The energy threshold for this process is:
\begin{equation}
 E_{\rm th~p\gamma}={m_{\rm p}c^2\epsilon_{\rm th~p\gamma}}/2\epsilon=(5\times 10^4\,{\rm TeV})\,(T_4)^{-1}\,,
\label{thres1}
\end{equation}
where $m_{\rm p}$ is the proton mass and $\epsilon$ and $\epsilon_{\rm th~p\gamma}\approx 140$~MeV 
are the energy of the target photon in the laboratory and the hadron rest frames, respectively. The loss rate is given by:
\begin{equation}
t_{\rm p\gamma}\sim{10^{18}\over N_{\rm X}}\, {\rm s}\,,
\label{pgam1}
\end{equation}
where:
\begin{equation}
N_{\rm X}\approx {L\over4\pi\epsilon R^2c}\approx 2\times 10^{14}L_{38}T_4^{-1}R_{12}^{-2}\,{\rm cm}^{-3}\,.
\end{equation}
$L_{38}$ is the star luminosity $L_*/10^{38}$~erg~s$^{-1}$, $R_{12}$ is the distance to the star, 
and $R/10^{12}\, {\rm cm}$. Eq.~(\ref{pgam1}) can be rewritten as:
\begin{equation}
t_{\rm p\gamma}\sim 10^4L_{38}^{-1}T_4R_{12}^2\, {\rm s}\,.
\end{equation}
With this cooling timescale, the corresponding maximum energy is:
\begin{equation}
E_{\rm max}\sim 10^5 \eta^{-1} B_{\rm a, G} L_{38}^{-1} T_4R_{12}^2\,{\rm TeV}\,.
\label{emaxpm}
\end{equation}

As for pp interactions, the long cooling timescales of photo-meson production imply  that the maximum energy is in fact
limited by the accelerator size. We note that only in case a substantial part of the energy in accelerated protons were 
$>E_{\rm th~pm}\sim 10^4$~TeV, under an UV photon field, and the fastest protons escaped at the speed of light, photo-meson
production could reach efficiencies of $L_{\gamma}\sim 10^{-3}L_{\rm p}$\footnote{At the involved proton energies,  the
stellar wind can hardly confine the particles, thereby we adopt the speed of light as escape velocity.} for $E_{\rm p}>E_{\rm
>th~p\gamma}$. The energy transfer
efficiency for this process is $c_{\rm p\gamma}=0.1$. In the inner regions of the jet, close to the accretion disk and corona
photon fields, the radiation energy density may be high enough to get even higher efficiencies. The larger target photon 
energy would imply a reduced threshold energy. Nevertheless, the constraint imposed by 
Eq.~(\ref{maxelarm}) is quite restrictive and could prevent  hadrons from reaching energies $>E_{\rm th~p\gamma}$ for
reasonable $B_{\rm a}$ and $l_{\rm a}$ values. Namely, comparing Eqs.~(\ref{emaxpm}), (\ref{thres1}) and (\ref{maxelarm}), two  
conditions are obtained:
\begin{equation}
B_{\rm G}T_4l_{11}>2\times 10^3\,,
\label{cond1}
\end{equation}
for basic physical conditions; and
\begin{equation}
\eta<B_{\rm a,G}T_4^2R_{12}^2/2L_{38}\,.
\label{cond2}
\end{equation}
for the acceleration efficiency. 

As in the case of pp collisions, the  strong dilution of the photon field far
from its source prevents photo-meson production from being efficient outside the binary system.

{\bf Photo-disintegration}

If ultra relativistic 
heavy nuclei are present, they can suffer photon disintegration under the ambient photon field. The expression for the 
threshold energy is similar to that presented in Eq.~(\ref{thres1}):
\begin{equation}
 E_{\rm th~pd}={m_{\rm N}c^2\epsilon_{\rm th~pd}}/2\epsilon\,,
\label{thres}
\end{equation}
where $m_{\rm N}$ is the mass of the nucleus and $\epsilon_{\rm th~pd}=8$~MeV. Effectively, 
since $m_{\rm N}$ can be up to $\sim 100\,m_{\rm p}$, $E_{\rm th~pd}$ could be $>E_{\rm th~p\gamma}$.

The disintegration of the nuclei has as typical timescale:
\begin{equation}
t_{\rm pd}\sim 10^3L_{38}^{-1}T_4R_{12}^2\,{\rm s}\,.
\label{dis}
\end{equation}
For simplicity, we have assumed that the mass of the nucleus is its charge times the mass of the proton, 
which slightly overestimates the efficiency within
a factor of $\sim 2$. In addition, a slow dependence on the mass of the nuclei in the cross section has 
been neglected. Photo-disintegration would stop the acceleration of heavy nuclei at energies:
\begin{equation}
E_{\rm max}\sim 10^4\frac{q}{e} B_{\rm a, G} L_{38}^{-1}T_4R_{12}^2\,{\rm TeV}\,.
\label{dis}
\end{equation}
As in the case of pp collisions and photo-meson production, the accelerator size instead of photo-disintegration cooling 
will probably stop acceleration of heavy nuclei.
Taking into account that the energy transfer efficiency is $c_{\rm p\gamma}\sim 0.01$, and 
adopting the speed of light to derive a lower-limit for $t_{\rm esc}$,
photo-disintegration can yield in gamma-rays $L_{\gamma}\sim 10^{-3}L_{\rm N}$, 
where $L_{\rm N}$ is the energy stored in heavy nuclei with $E>E_{\rm th~pd}$.  
Hillas constraint allows energies $m_{\rm N}/m_{\rm p}$ times larger for nuclei than for protons, although 
as noted above 
$E_{\rm th~pd}$ can be $>E_{\rm th~p\gamma}$\footnote{Actually, similar constraints to those derived for photo-meson production 
regarding the interaction physical conditions and the acceleration efficiency apply here as well.}. Like in the case of photo-meson production, 
it is unclear whether enough energy can be in the form of ultra relativistic 
heavy nuclei above the threshold to produce a significant signal via photo-disintegration.

{\bf Photon-photon absorption and secondary radiation}

\begin{figure}[pb]
\centerline{\psfig{file=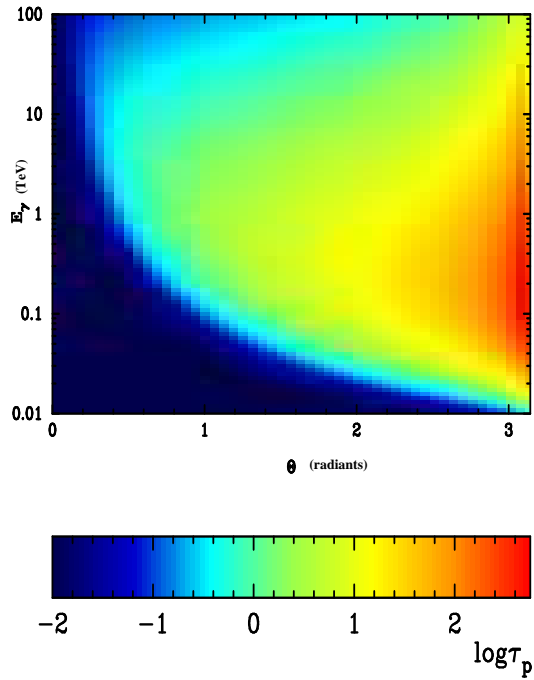,width=9cm, angle=270}}
\vspace*{8pt}
\caption{2-dimensional $\theta-E_{\rm gamma}$ map of the opacity coefficient. 
The adopted parameter values are $L_*=10^{39}$~erg~s$^{-1}$, $d_*=3\times 10^{12}$~cm and $kT_*=3$~eV. \label{absor}}
\end{figure}

\begin{figure}[pb]
\centerline{\psfig{file=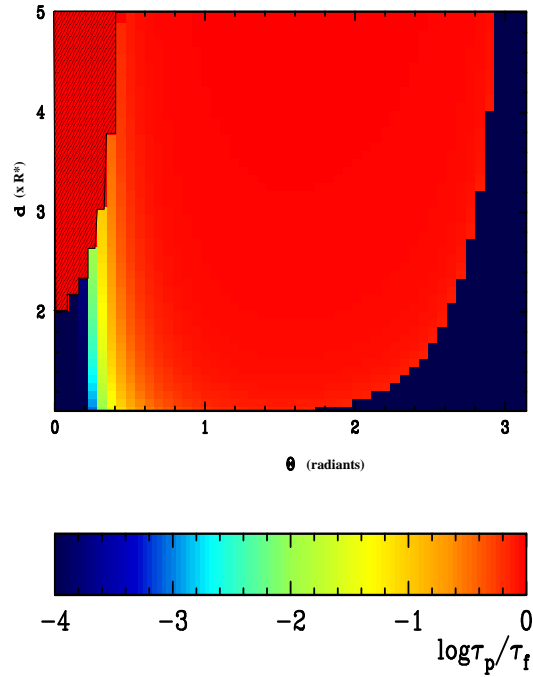,width=9cm, angle=270}}
\vspace*{8pt}
\caption{2-dimensional $\theta-d$ map of the ratio $\tau_{\rm p-l}/\tau_{\rm fin}$ for an incoming photon of 1~TeV. 
The adopted parameters are the same as those of Fig.~\ref{absor}, plus the stellar radius $R_*=7\times 10^{11}$~cm. 
The shaded area to the left of the plot corresponds to a region of opacities $\sim 0$. \label{fin}}
\end{figure}

The presence of the star, a powerful source of UV photons, is very relevant to understand TeV emission from compact sources.
A massive and hot star is an excellent provider of target photons for photon-photon absorption. The anisotropy of the stellar
photon field and the relative position of the system with respect to the observer makes the exploration of the observational
impact of photon-photon absorption a non-trivial subject. In Fig.~\ref{absor}, we show a 2-dimensional map of the
opacity coefficient ($\tau$) for different energies of the incoming photon ($E_{\gamma}$) emitted
near the compact object, and different directions starting
with the one pointing away from the star. Here, $\theta$ is the angle with the line joining the TeV emitter and the star. The
point-like approximation for the target photon field has been used, which implies that $\tau\propto 1/d$. The used parameter
values are $L_*=10^{39}$~erg~s$^{-1}$, $d_*=3\times 10^{12}$~cm and $kT_*=3$~eV. It is worthy noting that the point-like
approximation for the target photon emitter works in a wide range of situations, but there are cases relevant for compact TeV
sources in which is necessary to account for the finite size of the star (see also Dubus \cite{dubus06}). In Fig.~\ref{fin},
we show a 2-dimensional map of the ratio $\tau_{\rm p}/\tau_{\rm f}$ (i.e. point-like versus finite size $\tau$) for
different distances to the star ($d$) and values of $\theta$. The radius of the star has been taken $R_*=7\times 10^{11}$~cm.
The energy of the incoming photon has been taken 1~TeV. The deep blue/black regions in this map represent the cases when the
point-like approximation gives almost no absorption, i.e. when the emitter points roughly away from the star, or when the
photons in the finite size approximation collide with the star surface giving infinite opacity. 

\begin{figure}[pb]
\centerline{\psfig{file=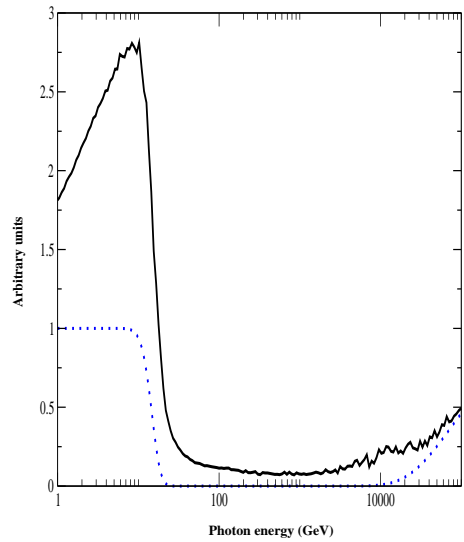,width=9cm, angle=270}}
\vspace*{8pt}
\caption{1-dimensional electromagnetic cascade. Note the difference 
of about a factor of 10 between the 1~TeV and the 10~GeV fluxes.
The adopted parameters are those of Fig.~\ref{absor}.
\label{casc}}
\end{figure}

The importance of cascading has been already stated above. It is worth now to study the possibility of cascade development in
the surroundings of massive and hot stars. There are two extreme situations. Either the magnetic field energy density is much
lower compared with the radiation energy density, and in such a case pure cascade will develop, or the magnetic field energy
density is a significant fraction, or above, the radiation one, and then the electron energy will be mostly released via
synchrotron radiation and single-scattering IC. If the magnetic field is low enough (and therefore suitable for cascading to
occur), very high-energy particles will be only slightly deflected before radiating their energy via IC. This allows us for a
1-dimensional approach to compute cascading. For the same parameter values as those presented above, we have performed a
1-dimensional electromagnetic cascade simulation, the result of which is shown in Fig.~\ref{casc}. The injected particle 
spectrum was a power-law of photon index 2.

\begin{figure}[pb]
\centerline{\psfig{file=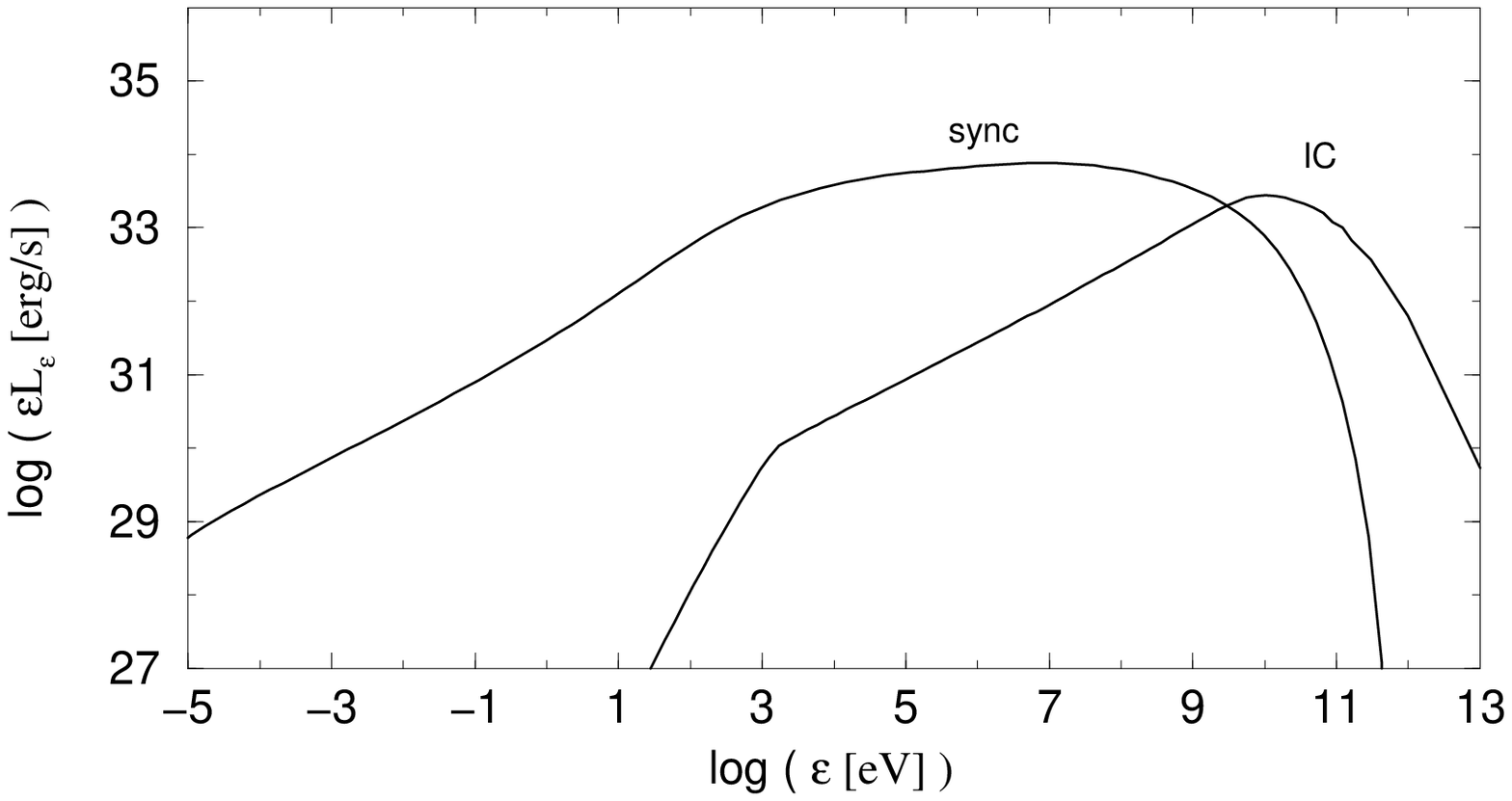,width=9cm, angle=0}}
\vspace*{8pt}
\caption{Computed SED of the synchrotron and IC emission from the pairs created in a compact
binary. The TeV emitter location is at $d=3\times 10^{12}$~cm. The adopted parameters are those of
Fig.~\ref{absor}, plus a stellar surface magnetic field of 100~G and a primary gamma-ray injection luminosity
of $3\times 10^{35}$~erg~s$^{-1}$.\label{sedsec}}
\end{figure}

To give an idea of the magnetic field importance, 
next formula shows the critical value of the magnetic field that allows cascading to occur for 
TeV electrons:
\begin{equation}
B_{\rm c}=10\,\frac{R_*}{R}\left(L_*\over 10^{39}\,{\rm erg~s}^{-1}\right)^{1/2}\,{\rm G}\,, 
\label{eq:b_crit}
\end{equation}
In fact, $B_{\rm c}$ is valid for the 1-dimensional cascade case. If electrons suffer strong deflection in the ambient
magnetic field, the longer time required by particles to escape the region of dense photon field will increase the
synchrotron outcome with respect to the IC one. This happens since photons convert to electrons more times before escaping,
releasing at the end more energy in form of synchrotron and low energy IC radiation. From all this, and the fact that
magnetic fields of hundreds of G are typical in the surface of OB stars, we conclude that effective electromagnetic cascading
is unlikely in the environment of high-mass systems. For the same conditions as those taken for Fig. 6, plus a stellar surface
magnetic field of 100~G and a primary gamma-ray injection luminosity of $3\times 10^{35}$~erg~s$^{-1}$, we show the secondary
pair spectral energy distribution (SED) in Fig.~\ref{sedsec}. We note the moderately high radio and X-ray luminosities (see
also Bosch-Ramon et al. \cite{bosch08}).

\subsection{Modeling non-thermal emission in microquasars: the case of Cygnus~X-1}\label{ls}

In Sect.~\ref{nonther}, we have reviewed the main radiative processes that could take place under typical conditions in
microquasars. After roughly estimating the efficiency of different leptonic and hadronic processes, it is our aim now to
focus on the mechanism that, to our understanding, is the most likely to produce the VHE radiation observed in some of these
sources. From the point of view of efficiency, IC scattering is clearly a good candidate in this sense. In addition, the fact
that a hot and massive star is present in all these systems detected at TeV energies makes IC scattering very
attractive\footnote{ It does not discard low-mass microquasars as VHE emitters. Actually, from the observational point of
view, hints of TeV emission from GRS~1915$-$105 were found using imaging Cherenkov techniques by HEGRA (Aharonian
\& Heinzelmann \cite{aharonian98b}) and Whipple (Rovero et al. \cite{rovero02}). In addition, a number of theoretical works
have proposed these sources as VHE emitters (e.g. Atoyan \& Aharonian \cite{atoyan99}; Bosch-Ramon et al. \cite{bosch06b};
Romero et al. \cite{romero08}).}. Finally, the pattern of variability of the spectrum and flux along the orbit found in
LS~5039 can be, together with photon-photon absorption, explained due to the angular dependence of the cross section of IC
scattering and the anisotropic nature of the stellar photon field (e.g. Khangulyan \& Aharonian \cite{khangulyan05};
Khangulyan et al. \cite{khangulyan07}). 

\begin{figure}[pb]
\centerline{\psfig{file=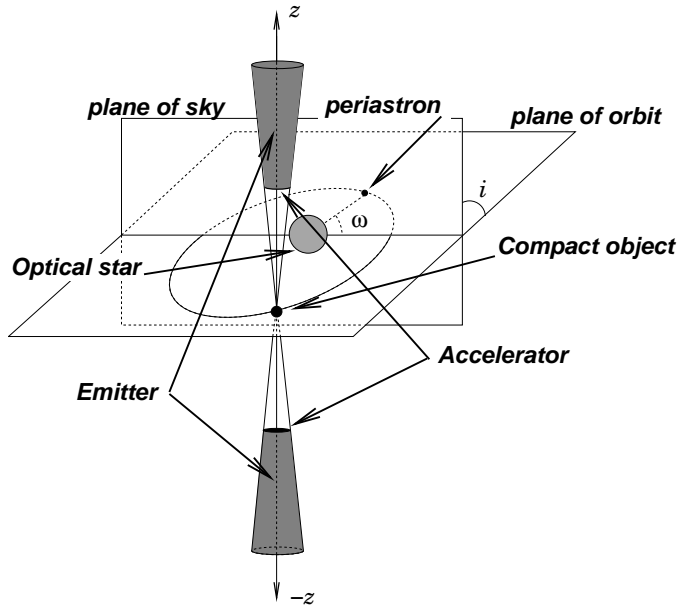,width=9cm}}
\vspace*{8pt}
\caption{Schema of the system. \label{sq}}
\end{figure}

To carry out a simple but thorough treatment of the processes relevant to VHE emission in microquasars, we have simplified
the scenario as much as possible. We show a schema of this scenario in Fig.~\ref{sq}. We consider the jet as a weakly
relativistic convective flow without going into the details of its physical nature, i.e. whether it is a magneto- or a
hydrodynamical plasma, or even a pointing flux dominated flow. This outflow is ejected perpendicular to the orbital plane
bearing a disordered magnetic field attached to it. Efficient acceleration of non-thermal particles can occur at a certain
location in the jet ($Z_0$ from the compact object, $d_0$ from the star); such acceleration regions are treated as point-like,
and their position in the jet, $Z_0$, could change. For simplicity, we adopt a model in which acceleration takes place only
in one point.

Because of the strong uncertainties affecting the physics of microquasar jets, we do not treat but very generally the
acceleration processes that may take place in there. The injection electron spectrum adopted here is phenomenological: a
power law of exponent -2 plus an exponential high-energy cutoff. The accelerator sets up the initial conditions of the
emitter, which is treated as a 1-dimensional region where particles are transported by jet convection. Adiabatic losses have
been neglected here. The magnetic field depends on $Z$ like $B=B_{\rm a}(Z_0/Z)$. To model the non-thermal processes included
in our scenario, it is necessary to consider the presence of different ingredients: the magnetic field, possibly present
radiation fields, and the material in which the emitting particles may be embedded. As discussed above, we will deal here
with synchrotron and stellar IC emission, since we restrict ourselves to high mass systems and to the most relevant processes
to produce or affect VHE radiation. Nevertheless, other mechanisms cannot be discarded. We take into account the impact of
photon photon absorption on the VHE spectrum. We recall the importance of the geometry of IC scattering and photon photon
absorption for the spectra and lightcurves when the star is the main source of target photons. For further mathematical
details of the model, we refer to Khangulyan et al. \cite{khangulyan07}.

In the following, we apply this model to the microquasar Cygnus~X-1 (Sect.~\ref{mQc}), and to the other two X-ray binaries
with extended radio emission, LS~5039 and LS~I~+61~303 (Sect.~\ref{lslsi}).

\subsubsection{Application to Cygnus~X-1}\label{mQc}

Cygnus~X-1 is a HMXB of uncontroverted accreting nature with relativistic radio jets (Stirling et al. \cite{stirling01}). The
compact object is a black-hole of $\sim 20$~M$_{\odot}$, the primary is an O9.7Iab star of $\sim 40$~M$_{\odot}$, and the
system is located at $\approx 2.1$~kpc (Ziolkowski \cite{ziolkowski05}). The orbit is thought to be circular, with an
inclination $i\sim 30^{\circ}$, orbital radius of $\approx 3\times 10^{12}$~cm and period of 5.6~days (Gies \& Bolton
\cite{gies86}). The primary star presents a luminosity of $\approx 10^{39}$~erg~s$^{-1}$ and a temperature of $\approx
30000$~K. An evidence of detection from this source at the 4.1~$\sigma$ level has been found by MAGIC in the TeV range during
a transient event that may be correlated with the hard X-ray behavior (Albert et al. \cite{albert07}). At present, the origin
of the VHE emission is unclear. In the context of the hadronic scenario, Romero et al. \cite{romero03} and Orellana et al.
\cite{orellana07} computed the  emission for a microquasar with the characteristics of Cygnus~X-1. Regardless the hadronic or
leptonic origin of the gamma-ray radiation, electromagnetic cascading and/or secondary synchrotron emission should occur in
the system (e.g. Bednarek \cite{bednarek07}, Orellana et al. \cite{orellana07}, Bosch-Ramon et al. \cite{bosch08}).

{\bf On the spectral energy distribution and orbital variability}

\begin{figure}[pb]
\centerline{\psfig{file=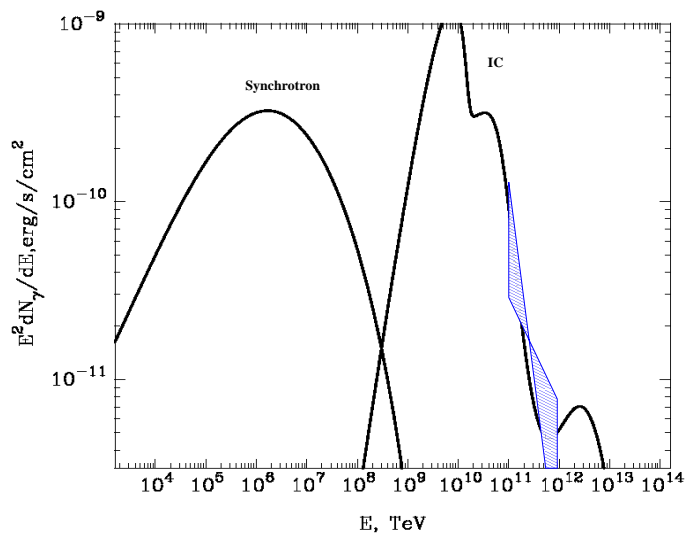,width=9cm, angle=0}}
\vspace*{8pt}
\caption{Computed synchrotron and IC SED for Cygnus~X-1, together with the observed VHE SED, is shown. The accelerator/emitter
location has been set to $Z_0=3\times 10^{12}$~cm. The total contribution from the jet and the counter-jet is
shown. 
\label{sedcyg}}
\end{figure}

It is worthy to see the aspect of the multiwavelength non-thermal emission inferred from the VHE radiation detected from the
source. This is shown in Fig.~\ref{sedcyg}. We have located the accelerator/emitter at $Z_0=3\times 10^{12}$~cm. The
magnetic field is $\sim$~G. The synchrotron X-ray flux is well below the observed level, of likely thermal (comptonized)
origin (Sunyaev \& Truemper \cite{sunyaev79}). This SED has been calculated for an emitter located in the borders of the
system, in which opacities are moderate. However, deeper, closer to the compact object, the attenuation factor grows orders
of magnitude.

For Cygnus X-1, the spectrum of the emission from $\sim 0.1-1$~TeV electrons is available for a narrow orbital phase range
around $\phi=0.9$, i.e. this is not averaged for wide ranges of the orbit. In addition, these electrons have the shortest
lifetimes under IC cooling, implying that they radiate most of their energy before propagating significantly. These two facts
mean that the observed radiation was produced under similar conditions. In addition, the stellar photon field is very dense
in the compact object surroundings, implying very large opacities in almost all the directions. All this, plus the known
orbital observer/system geometry at the observation phase, allows us, and makes interesting, to estimate the absorbed energy
via photon-photon absorption depending on the distance between the emitter and the compact object. The calculation of the
energy processed in this system via photon-photon absorption can give a model-independent constraint on the the location of
the TeV emitter. 

\begin{figure}[pb]
\centerline{\psfig{file=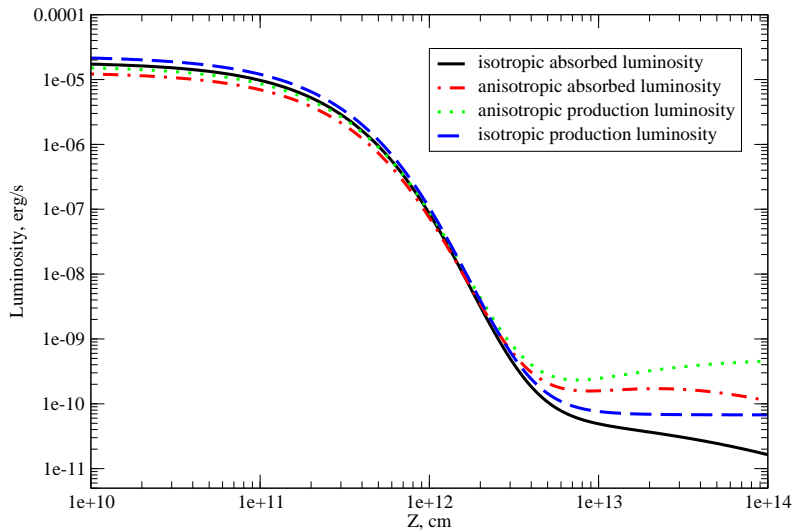,width=9cm, angle=270}}
\vspace*{8pt}
\caption{Luminosity divided by $4\pi d_{\odot}^2$ of the secondary pairs injected in the system with the emitter at different
distances from the compact object. This is computed calculating first the deabsorbed VHE spectrum and flux given the distance
to the star and the observed spectrum and flux. The case in which the radiation process is isotropic, like proton-proton
collisions may be (solid line), and the case in which the radiation process is IC (dot-dashed line) -strongly anisotropic-,
are shown. The production curves for both cases are also shown.
\label{abscyg}}
\end{figure}

In Fig.~\ref{abscyg}, we show the luminosity divided by $4\pi d_{\odot}^2$ ($d_{\odot}$: distance to the Earth) of the
secondary pairs injected in the system as a function of the distance to the compact object. This is computed calculating
first the deabsorbed VHE spectrum and flux given the distance to the star and the observed spectrum and flux (Albert et al.
\cite{albert07}). In our calculations we have not taken into account the effect of cascading, although the likely
moderate-to-high magnetic fields present in massive binary systems would imply a dominant synchrotron channel suppressing
cascading. Since in our
context a significant fraction of this energy rate is released via synchrotron emission, and the X-ray/soft gamma-ray
luminosities are typically of $\sim 10^{37}$~erg~s$^{-1}$ (e.g. McConnell et al. \cite{mcconnell02}), the emitter location
can be set, conservatively, to at least few times $10^{12}$~cm (see also Bosch-Ramon et al. \cite{bosch08b}). This is a
strong indication that the TeV emitter is located in a jet far away from the compact object. It is relevant also to note that
the secondary synchrotron radiation may also explain the observed soft gamma-ray fluxes (e.g. McConnell et al.
\cite{mcconnell02}). Another interesting result is the absorbed luminosity at distances $\sim 10^{13}$~cm, indicating that
there can be significant injection of relativistic pairs at distances where their radio emission may be resolved using VLBI
techniques (see also Bosch-Ramon et al. \cite{bosch08}). 

Regarding the time evolution of the observed radiation, the flaring nature of the VHE emission points to intrinsic changes of
the emitter properties, instead of geometric effects or target density variations, as the origin of the variability.

\subsection{TeV binaries with extended radio emission: LS~5039 and LS~I~+61~303}\label{lslsi}

\subsubsection{LS~5039}\label{ls}

LS~5039 is an HMXB (Motch et al. \cite{motch97}) located at 2.5$\pm 0.5$~kpc (Casares et al. \cite{casares05}). The source
presents radio jets (Paredes et al. \cite{paredes00,paredes02}), shows X-ray variability possibly associated to the orbital
motion (Bosch-Ramon et al. \cite{bosch05}), and has been detected at very high-energy (VHE) gamma-rays (Aharonian et al.
\cite{aharonian05}), which virtually confirms its association with a CGRO/EGRET source (Paredes et al. \cite{paredes00}).
Interestingly, the TeV emission varies with the orbital period (Aharonian et al. \cite{aharonian06}). The most recent orbital
ephemeris of the system were obtained by Casares et al. \cite{casares05}. The orbital period is $3.9060$~days, the
eccentricity is moderate $e=0.35\pm 0.04$, and the inclination angle is not well constrained, $i\approx
15^{\circ}$--$60^{\circ}$. The orbital semi-major axis of the system is $\approx 2\times 10^{12}$~cm. The nature of the
compact object is still uncertain. Casares et. al. \cite{casares05} suggested that it may be a black hole\footnote{Assuming
pseudo-synchronization of the orbit.}, although there is an on-going debate on this issue, and some authors have proposed
that LS~5039 is in fact a young non-accreting pulsar (see, e.g., Martocchia et al. \cite{martocchia05}; see also Dubus
\cite{dubus06b}). In this regard, the strongest argument would the lack of accretion features in the X-ray spectrum of
LS~5039. 

\begin{figure}[pb]
\centerline{\psfig{file=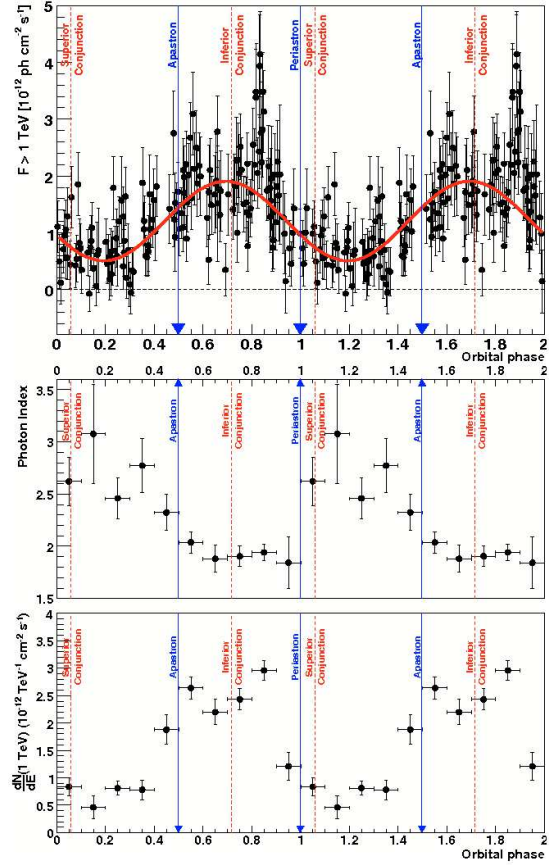,width=9cm, angle=0}}
\vspace*{8pt}
\caption{At the top, the lightcurve of the photon flux above 1~TeV of LS~5039 presented by Aharonian
et al. (2006). Two full phase periods are shown for clarity. The blue solid arrows depict periastron
and apastron. The thin red dashed lines represent the superior and inferior conjunctions of the
compact object, and the thick red dashed line depicts the Lomb-Scargle Sine coefficients for the
period giving the highest Lomb-Scargle power (see Aharonian et al. 2006 for details). At the
middle, the fitted pure power-law photon index (for energies 0.2 to 5~TeV) versus phase interval of
width $\phi=0.1$ is presented. At the bottom, the power-law normalization (at 1~TeV) versus. phase
interval of width $\Delta\phi=0.1$ is shown. \label{lslc}}
\end{figure}

\begin{figure}[pb]
\centerline{\psfig{file=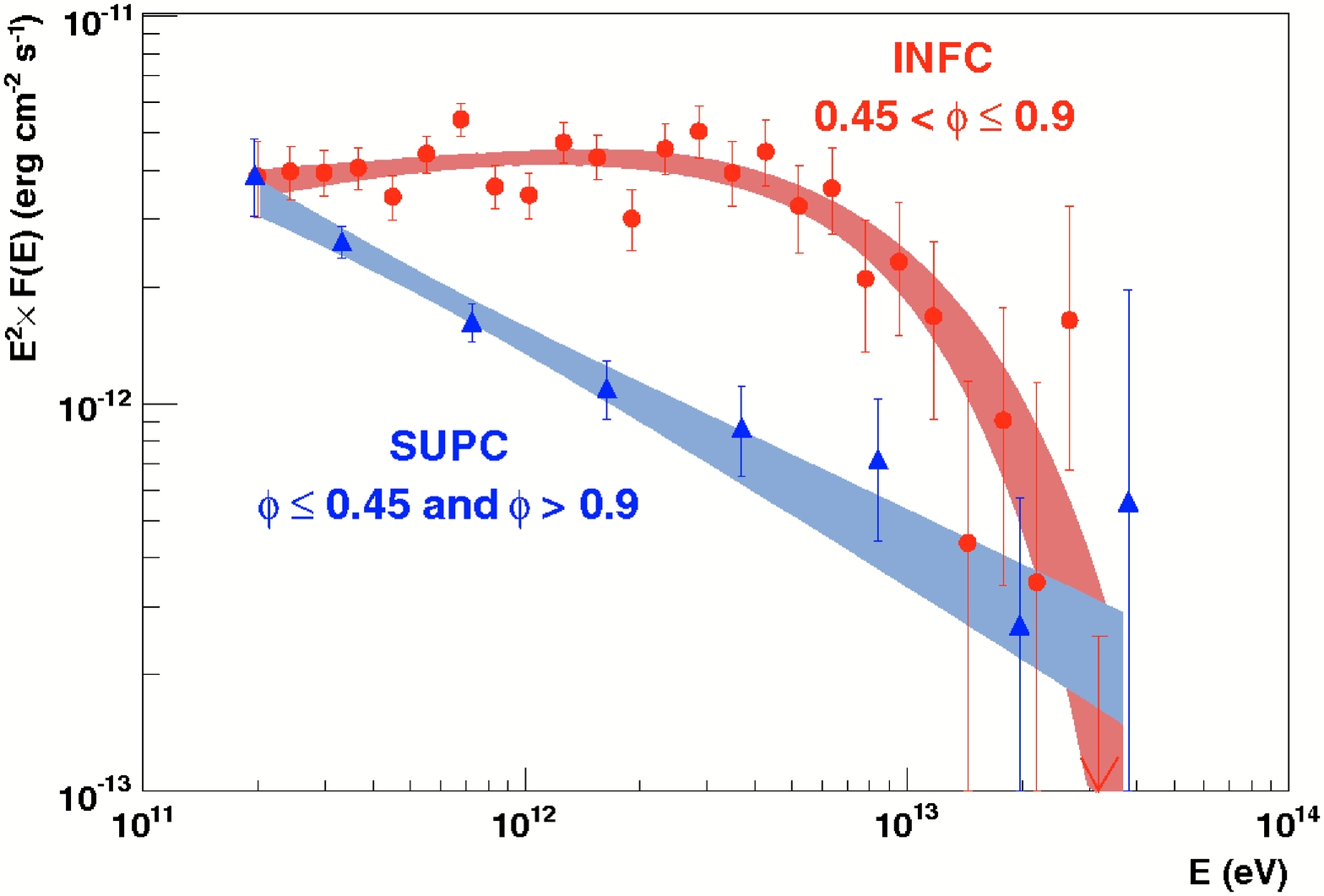,width=9cm, angle=0}}
\vspace*{8pt}
\caption{Spectra of LS~5039 for the two broad orbital phase intervals $0.45<\phi\le 0.9$ (red circles),
and  $\phi\le 0.45$ and  $\phi>0.9$ (blue triangles), from Aharonian et al. (2006). The shaded regions
represent the 1~$\sigma$ 
confidence bands on the fitted functions (see Aharonian et al. 2006 for details). \label{lssed}}
\end{figure}

The radio emission in LS~5039 is of non-thermal origin, slightly variable at month--year timescales ($\sim 20$--$30$\%) and
extended -$\sim$1~--~100~milliarcseconds (mas)- (Mart\'i et al. \cite{marti98}; Rib\'o et al. \cite{ribo99}; Paredes et al.
\cite{paredes00,paredes02}, Rib\'o et~al.  \cite{ribo08}; for an exhaustive study of this source, see Rib\'o \cite{ribo02}),
the $\sim 60$--$80$\% being produced within a core of $\sim$~mas. The radio emission, when observed at VLA angular
resolution, appears unresolved, and optically thin (Mart\'i et al. \cite{marti98}).

In the X-rays, the source shows flux variations by a factor of $\sim 2$ peaking smoothly around phase 0.8 and more sharply at
other phases (Bosch-Ramon et al. \cite{bosch05}). These peaks were apparently accompanied by spectral hardening. At phases
$\sim 0.8$, higher and harder emission than in the rest of the orbit has been also observed at TeV energies (Aharonian et al.
\cite{aharonian06}), and simultaneous Chandra observations have apparently shown a similar behavior in X-rays (Horns et al.
\cite{horns06}; Bosch-Ramon et al. \cite{bosch07b}).

At VHE, the flux and the photon index change periodically, by a factor of $\sim 4$ the former and between $\sim 1.9$--$3.1$
the latter, with the spectrum hardening when flux increases (Aharonian et al. \cite{aharonian06}). The observed lightcurve
and VHE SED are shown in Figs.~\ref{lslc} and \ref{lssed}, respectively. The maximum of the emission takes place around phase
0.8, similarly as it seems to happen at X-rays. Also, sudden increases/hardening in the flux and spectrum on hour timescales
could have been detected at phase $\sim$0.85 (de Naurois et al. \cite{denaurois06}), being similar to what has been mentioned
above concerning sharp peaks in X-rays. The TeV emission from LS~5039 has been studied by several authors in the microquasar
(e.g. Paredes et al. \cite{paredes06}) and the pulsar context (e.g. Dubus et al. \cite{dubus07}; Sierpowska-Bartosik \&
Torres \cite{sierpowska07,sierpowska08}). A more general approach to study the VHE emission from LS~5039 has been recently carried out by
Khangulyan et al. \cite{khangulyan07}.

In the following, we present the multiwavelength SED and the orbital VHE lightcurve computed with the model sketched above
for LS~5039. The values given to the free parameters of the model are chosen with the intention to show that a simple
leptonic model can roughly explain the observed features of the emission taking into account the TeV data and the
phenomenology of the source. At this stage, no electromagnetic cascading should be introduced.

{\bf On the spectral energy distribution and orbital variability}

\begin{figure}[pb]
\centerline{\psfig{file=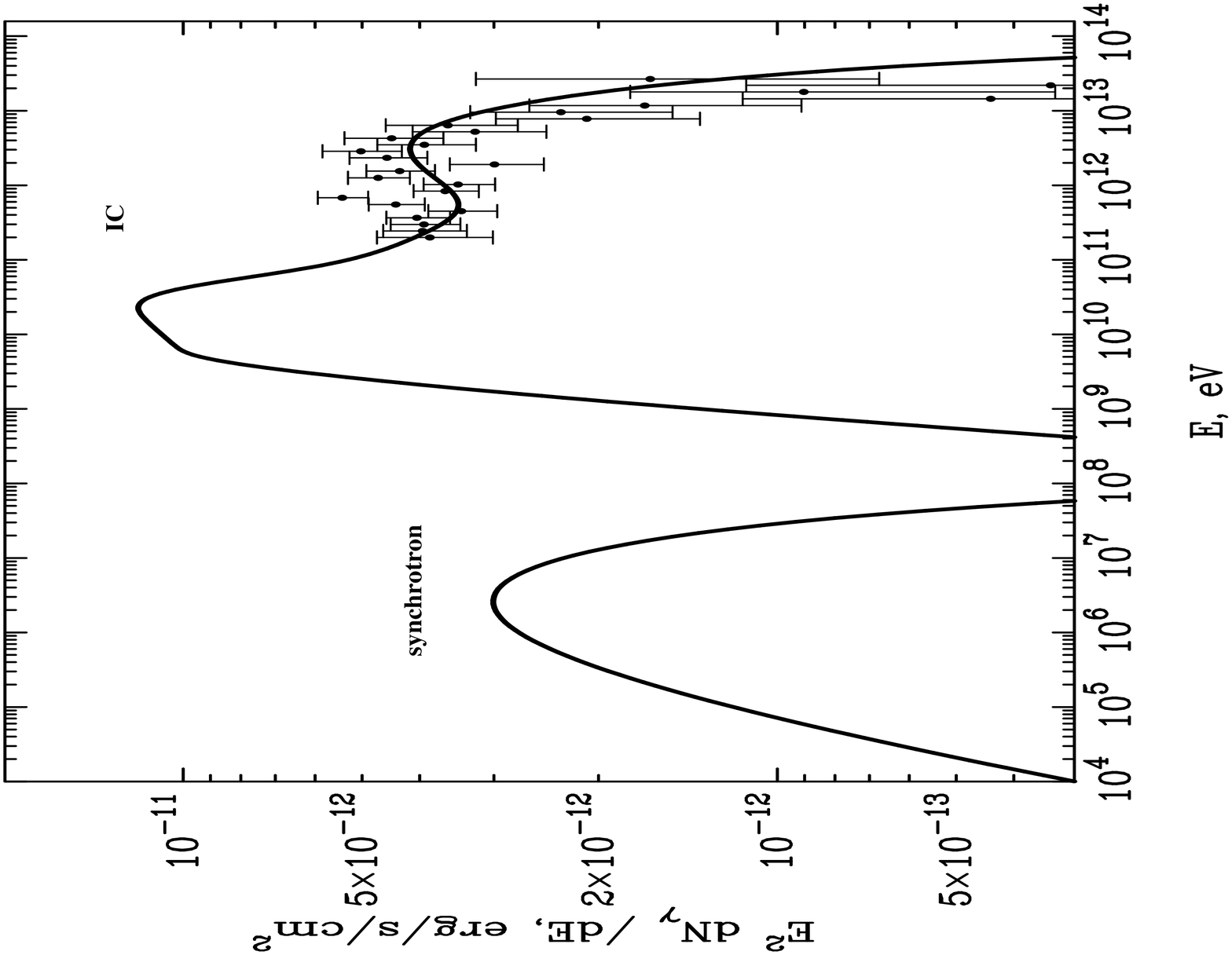,width=9cm, angle=270}}
\vspace*{8pt}
\caption{
Computed synchrotron and IC SED of LS~5039 averaged over the orbital phase interval $0.9>\phi>0.45$. The HESS data points are
also shown (Aharonian et al. 2006). The adopted parameter values are $V_{\rm adv}=0.1\,c$, $Z_0=10^{12}$~cm, $i=25^{\circ}$,
$\eta=10$, $B=0.05$~G,  and $L_{\rm e}=1.5\times 10^{35}$~erg~s$^{-1}$. The total contribution from the jet and the counter-jet is
shown. 
\label{sedls}}
\end{figure}

\begin{figure}[pb]
\centerline{\psfig{file=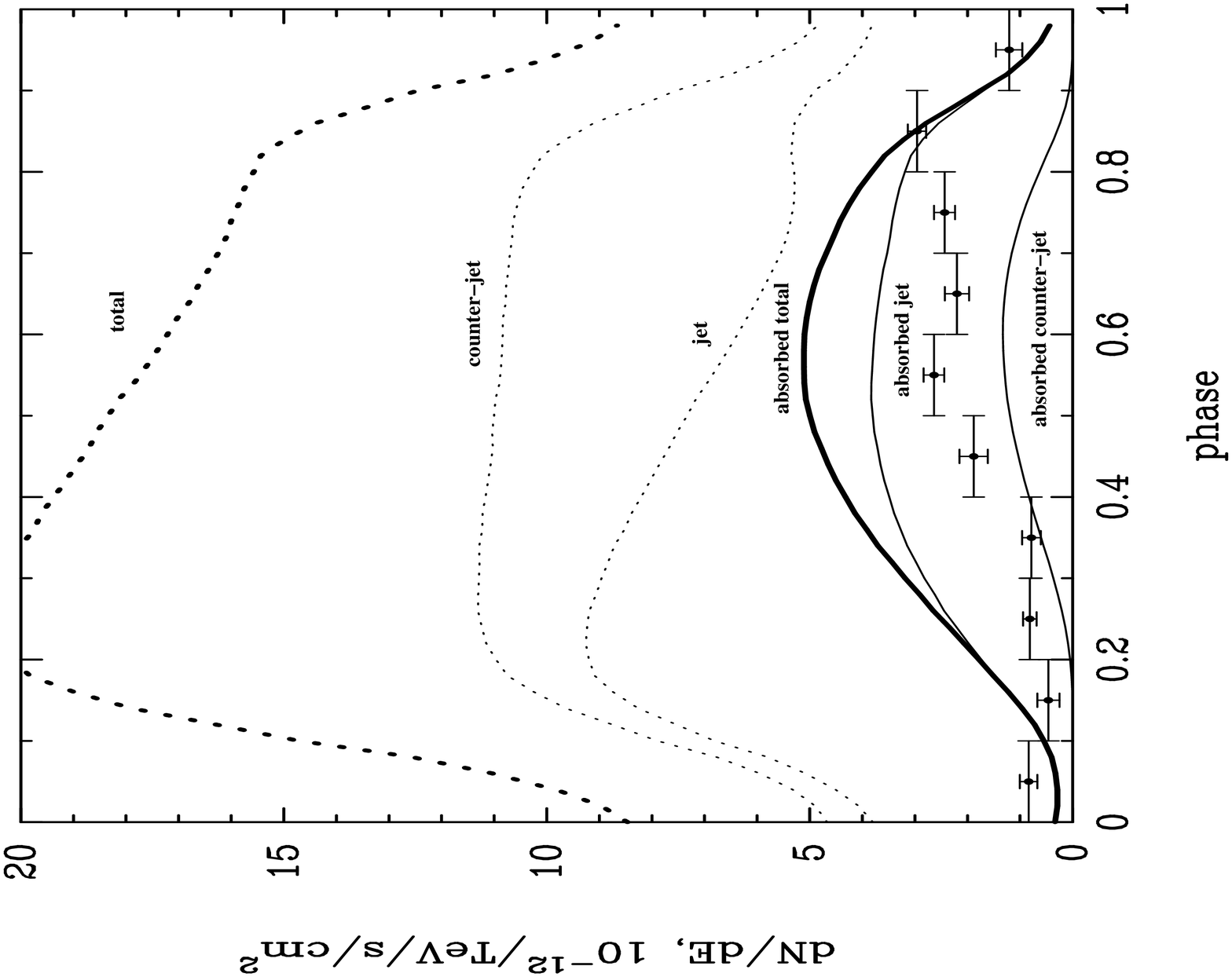,width=9cm, angle=270}}
\vspace*{8pt}
\caption{Computed IC orbital lightcurve of the differential photon flux at 1~TeV for LS~5039.
The different components, jet, counter-jet and summation of both, are labelled. The adopted
parameters are the same as those adopted in Fig.~\ref{sedls}. \label{lcls}}
\end{figure}

In Fig.~\ref{sedls}, for illustrative purposes, we show the synchrotron and IC SED for LS~5039 averaged over the orbital
phase interval $0.9>\phi>0.45$, corresponding to the inferior conjunction of the compact object ($\phi=0.72$). We use this
format for a better comparison with the spectral results from observations (Aharonian et al. \cite{aharonian06}), shown also
in the figure. For the very high energy spectrum in superior conjunction, see Khangulyan et al. \cite{khangulyan07}, fig.~23;
at that phase interval, the synchrotron emission will be similar to that at inferior conjunction. In Fig.~\ref{lcls}, the VHE
lightcurve along the orbit is shown. The adopted parameter values in both figures are the following: $V_{\rm adv}=0.1\,c$,
$Z_0=10^{12}$~cm, $i=25^{\circ}$, $\eta=10$ and $B=0.05$~G. The exact values of $B$ and $V_{\rm adv}$ are actually no
crucial. In the case of $B$, its value just needs to fulfill acceleration constraints (i.e. to allow electrons to reach high
enough energies to explain the TeV data), and be low enough such that VHE electrons will lose most of their energy via IC in
the KN regime. Regarding $V_{\rm adv}$, moderately different values will render similar results. To reach the observed
emission levels, the total injected luminosity in relativistic electrons is $L_{\rm e}\approx 1.5\times
10^{35}$~erg~s$^{-1}$. We note that we have computed the contribution from both the jet and the counter-jet and summed them
up. Photon-photon absorption is taken into account. Remarkably, the synchrotron part of the spectrum is well below the
observed fluxes, indicating that the dominant X-rays in this source likely come from a different emitting region. This region
should be nevertheless physically connected to the TeV emitter, given the similar behavior in time of both the X-ray and the
TeV radiation. 

Regarding variability, we show in Fig.~\ref{lcls} the orbital VHE lightcurve obtained for the same parameter choice adopted
to compute the SED. Although the matching with observations is not perfect, it is necessary to state that the lightcurve
provided in Aharonian et al. \cite{aharonian06} is constructed in bins with relatively small statistical significance, the
spectra are simplified to pure power laws (and they could be more complicated than just a power law, as shown in
Fig.~\ref{sedls}), and includes data points from observations of very different epochs. In any case, just playing with
propagation effects, IC scattering and photon photon absorption, one can already obtain very different lightcurves.

For an extensive discussion on the role of the different parameters present in the used model, we refer to Khangulyan et al.
\cite{khangulyan07}. Here our purpose is just to note that the leptonic scenario, accounting for changes in the interaction
geometry due to the orbital motion of the system, can reproduce quite nicely the observed spectra for different phases. It is
worthy mentioning that, to reach the observed maximum photon energies, $Z_0$ is to be $\ge 10^{12}$~cm unless the
acceleration efficiency is close to the maximum one,  i.e. $\eta<10$\footnote{In the case of a deep emitter, the magnetic
field in the system should be well below several G, as noted by Bosch-Ramon et al. \cite{bosch08b}.}. In addition, the
magnetic field in the accelerator/emitter is restricted to a relatively narrow band around 0.1~G.

\subsubsection{LS~I~+61~303}\label{lsi}

LS~I~+61~303 is a quite eccentric ($e=0.72$) HMXB, located at $\approx 2$~kpc, with an orbital semi-major axis of $\approx
6\times 10^{12}$~cm, a Be primary star of luminosity $\approx 10^{38}$~erg~s$^{-1}$ and temperature $\approx 28000$~K, and an
orbital period of 26.4960~days (Hutchings \& Crampton \cite{hutchings81}; Gregory \cite{gregory02}; Casares et~al.
\cite{casares05b}). The inclination of the system is not well constraint, being in the range $i=15-60^{\circ}$. Massi et al.
(\cite{massi01,massi04}) detected radio jets, with apparently relativistic motion, at $\sim 100$~milliarcsecond scales and
classified LS~I~+61~303 as a microquasar. Otherwise, the source does not show signatures of accretion in the X-rays (e.g.
Sidoli et al. \cite{sidoli06}; Paredes et al. \cite{paredes07}; and references therein), which are of likely non-thermal
origin. This fact, altogether with other arguments -mainly related to the extended radio emission morphology (see Dhawan et
al. \cite{dhawan06})- have let several authors to put forward a non-accreting pulsar as the compact object in the system
(e.g. Dubus \cite{dubus06b}; Chernyakova et al. \cite{chernyakova06}; Dhawan et al. \cite{dhawan06}), which had been proposed
for the first time by Maraschi \& Treves \cite{maraschi81}. Nevertheless, the microquasar scenario cannot be discarded, and
it may indeed explain some of the inferred properties of LS~I~+61~303 better than the pulsar scenario (e.g. Romero et al.
\cite{romero07}). In fact, the apparently slow and collimated radio structures detected in this source may be rather
difficult to explain in the colliding wind context, in which very fast motions of the shocked pulsar wind could be expected
(see Bogovalov et al. \cite{bogovalov08}). We recall that LS~I~+61~303 has been detected in the TeV range by MAGIC (Albert et
al. \cite{albert06}) with the maximum around the phase 0.6, being not detected during periastron passage, at phase 0.23
(Casares et al. \cite{casares05b}). This source had been also detected by EGRET (Kniffen et al. \cite{kniffen97}). It is
worthy noting that the lightcurves in the radio, X-rays, and high-energy and very high-energy gamma-rays look somewhat
similar (see fig.~3 in Chernyakova et al. \cite{chernyakova06} and references therein). Several authors have adopted
different frameworks and mechanisms to explain the high energy radiation from this source (in the microquasar framework: e.g.
Romero et~al. \cite{romero05b} -hadronic-,  Bosch-Ramon et al. \cite{bosch06c} -leptonic-, Bednarek \cite{bednarek06}
-cascading-; in the pulsar framework: e.g. Leahy \cite{leahy04}, Dubus \cite{dubus06b} -leptonic-, Chernyakova et al.
\cite{chernyakova06} -hadronic-; regarding neutrino detectability, see Christiansen et al. \cite{christiansen06} and Torres
\& Halzen \cite{torres07}).

{\bf On the spectral energy distribution and orbital variability}

\begin{figure}[pb]
\centerline{\psfig{file=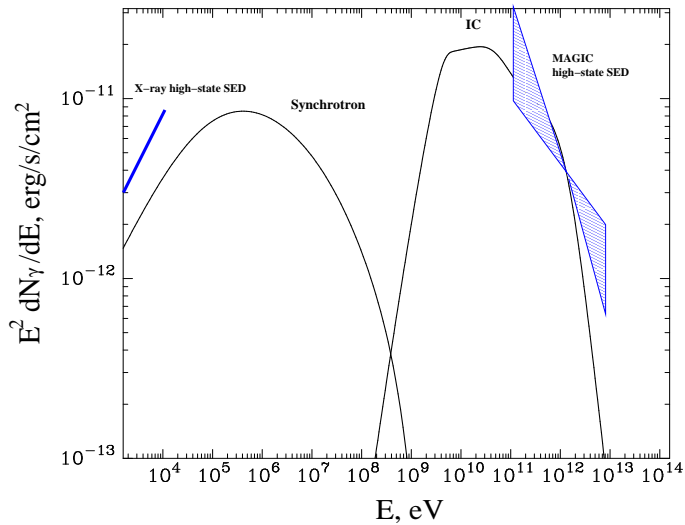,width=9cm, angle=0}}
\vspace*{8pt}
\caption{The Computed Synchrotron and IC SED for LS~I~+61~303 at phase $\phi=0.6$, 
together with the observed VHE SED, is shown. The adopted parameter values are 
$V_{\rm adv}=0.1\,c$, $Z_0=10^{12}$~cm, $i=30^{\circ}$, $\eta=10$, $B=0.5$~G 
and $L_{\rm e}=1.5\times 10^{35}$~erg~s$^{-1}$. The total contribution from the jet and the counter-jet is
shown.
\label{sedlsi}}
\end{figure}

\begin{figure}[pb]
\centerline{\psfig{file=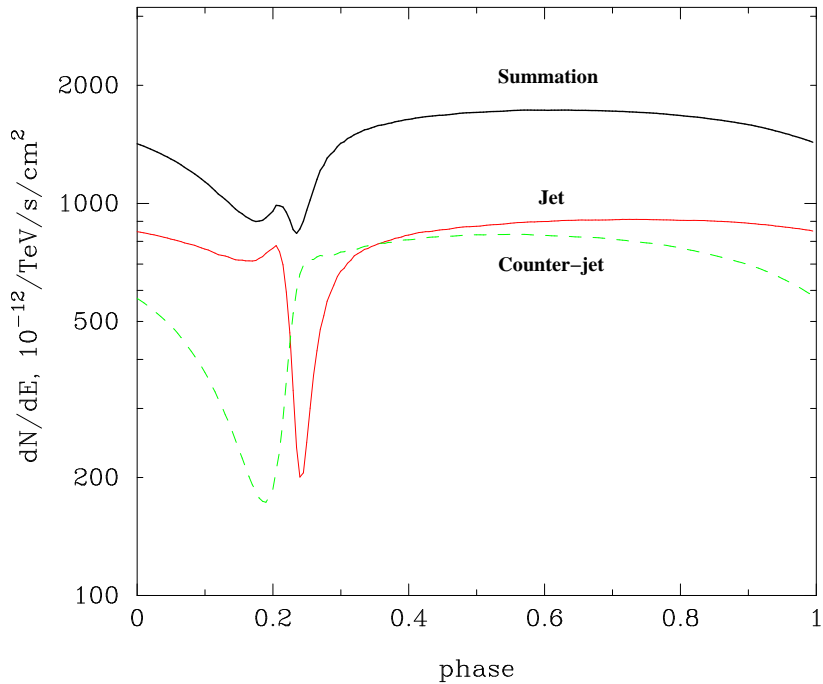,width=9cm, angle=270}}
\vspace*{8pt}
\caption{
The light-curve (differential photon flux) at 100~GeV for LS~I~+61~303. The jet and the counter-jet 
components are labelled. The adopted parameter values are the 
same as those in Fig.~\ref{sedlsi}.
\label{lclsi1}}
\end{figure}

\begin{figure}[pb]
\centerline{\psfig{file=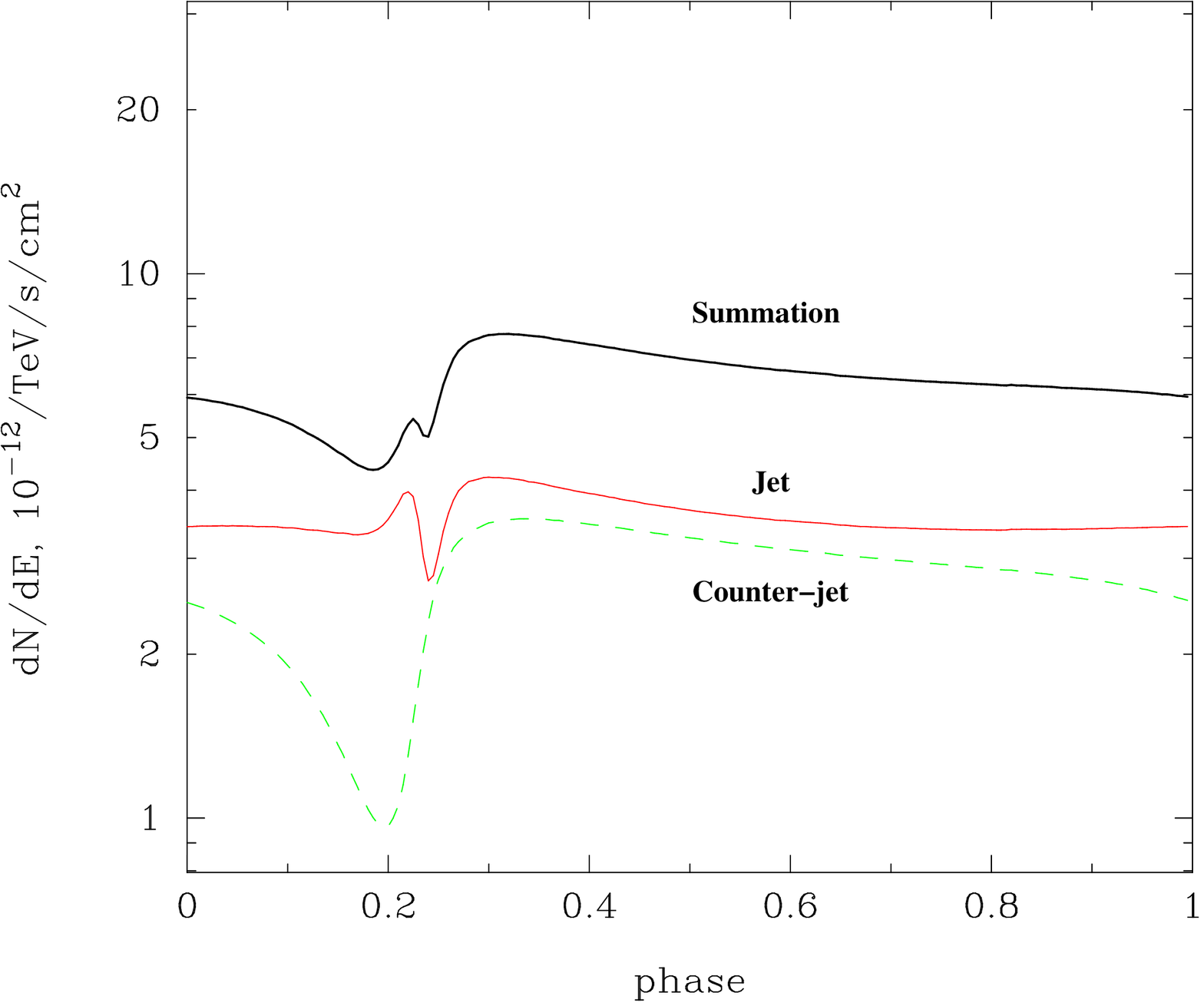,width=9cm, angle=270}}
\vspace*{8pt}
\caption{
The same as in Fig.~\ref{lclsi1} but at 1~TeV. 
\label{lclsi2}}
\end{figure}

\begin{figure}[pb]
\centerline{\psfig{file=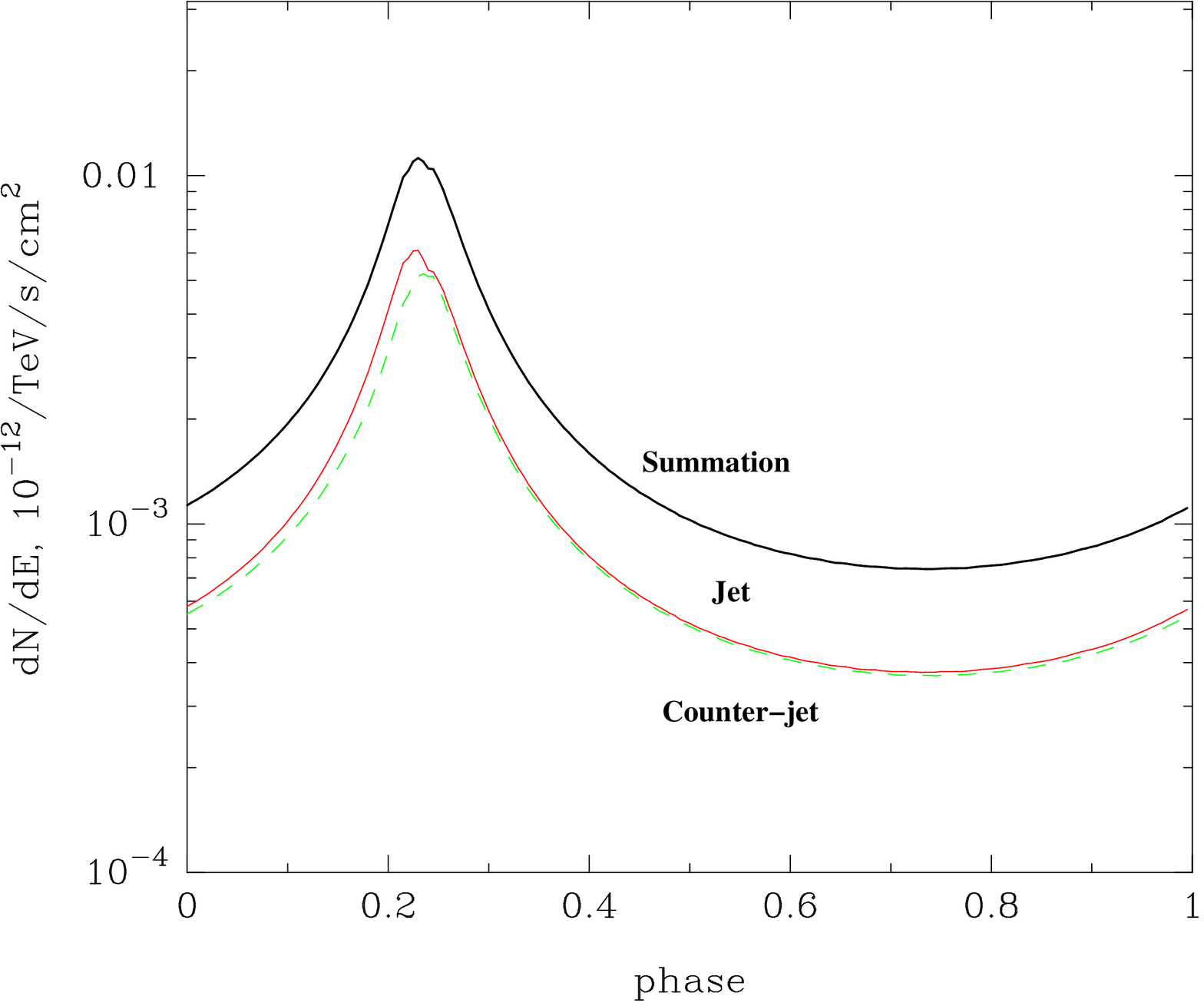,width=9cm, angle=270}}
\vspace*{8pt}
\caption{
The same as in Fig.~\ref{lclsi1} but at 10~TeV.
\label{lclsi3}}
\end{figure}

In Fig.~\ref{sedlsi}, we show the computed synchrotron and IC SED for LS~I~+61~303 at $\phi=0.6$. The adopted parameter
values are the following: $V_{\rm adv}=0.1\,c$, $Z_0=10^{12}$~cm, $i=30^{\circ}$, and $\eta=10$. In this case, the magnetic
field energy density could be larger than the radiation one compared to LS~5039, since the VHE spectrum is less hard, leaving
more room for synchrotron losses (which yield a softer spectrum). The value of $B$ obtained here to reproduce the observed
radiation is 0.5~G. The total adopted $L_{\rm e}$ is $1.5\times 10^{35}$~erg~s$^{-1}$. In LS~I~+61~303, the X-ray levels
would not be far from the observed ones. Still, for the adopted magnetic field, adjusted to explain the TeV spectrum, the
fluxes are still too low. Unlike LS~5039, neither the observer/binary system geometry nor the impact of photon-photon
absorption are enough when understanding the TeV variability, given the distance to, and the luminosity of, the primary star.
At the phase at which this emission is detected, the gamma-ray attenuation is low, and the IC scattering angle changes
relatively little. In Figs.~\ref{lclsi1}, \ref{lclsi2}, and \ref{lclsi3}, the differential photon flux lightcurves at 0.1, 1
and 10~TeV, respectively, are shown. In the observed lightcurve (Albert et al. \cite{albert06}), the maximum of the emission
is at phases (using the ephemeris of Casares et al. \cite{casares05b}; see also Grundstrom et al. \cite{grundstrom07})
different from those predicted by our IC/photon-photon absorption model. This fact, plus the similarity of the VHE lightcurve
with those at other wavelengths,v suggest intrinsic changes of the emitting region. Interestingly, the 10~TeV emission in our
scenario peaks roughly around periastron passage, due to the characteristics of the pair creation and IC cross sections. 

\section{Further comments}\label{disc}

We have discussed in this work the different processes that may take place in microquasars, and explored the possibility that
stellar IC scattering is behind the VHE radiation from three sources recently detected in the TeV range. We have shown also
the importance of photon-photon absorption, whose occurrence can be used to constrain the emitter properties. In this
section, we critically discuss three hot topics that are, in our view, very relevant for the kind of source discussed here.
The first point is related to the possibility, already mentioned and commented above, that a non-accreting pulsar may power
the non-thermal broadband emission in LS~I~+61~303 and LS~5039. The second point, affecting Cygnus~X-1 as well, is the role
played by the companion star concerning high-energy processes and the dynamics of the jet. Finally, some comments are made
regarding the prospects of the study of microquasars with the new/future VHE intrumentation.

\subsection{Evidences for a pulsar nature}

Although we do not consider here that LS~I~+61~303 and LS~5039 may harbor young non-accreting pulsars instead of an accreting
black-hole, this possibility cannot be discarded. At present, the strongest argument in favor of this is the lack of
accretion features in the X-ray spectrum. An {\it extension} of this argument is the fact that the timing properties of the
X-ray emission in these two sources, and even the radio/X-ray behavior, do not correspond to what is believed a microquasar
should be. Needless to say, this argument is phenomenological, and relies on a supposed regularity of the microquasar
behavior at X-rays. The same could be said regarding the radio behavior and the radio/X-ray connection.

Indeed, there is a whole set of phenomenological studies, as already mentioned in Sect.~\ref{genst}, majorly grounded on
observations but also to some extent on theoretical models, which try to give an unified microquasar picture. LS~5039 and
LS~I~+61~303 do not fit in such a picture regarding the points mentioned above. Nevertheless, most of the high-mass
microquasars, and to a lesser extent several low-mass microquasars, present some level of discrepancy from the standard
picture, and not all the sources may have occurring in them the same mechanisms as those comprehended in the standard
scenario. All this shows that such a picture or framework is a useful working hypothesis, but it cannot still be a
discrimination tool when trying to find out whether or not a source pertains to some class of objects. In the same line,
sometimes the morphology of the radio emission from these systems is used as an argument, again phenomenological, against
their microquasar nature. In this regard, it is claimed that the extended radio emission does not fit the standard picture of
microquasar jets. Nevertheless, it is worthy pointing out here that galactic compact jets are sometimes detected at the
resolution limits of radio VLBI interferometers. It makes therefore difficult to argue, based on solid observational grounds,
about how a microquasar jet should look like. From  the theoretical point of view, to define how a canonical microquasar jet
should be is presently not possible, since we lack a complete theory for jet formation and collimation, and there might be
more than one mechanism operating.

There are otherwise two good indicators of the presence of a pulsar in the system, i.e. the detection of radio and X-ray
pulsations, and the lack of strong X-ray radiation (as is the case) due to any form of accretion onto a possible neutron star
surface (in case the compact object were known to be a neutron star). To date, none of both questions has been answered.
Neither pulsations have been detected, nor the problem of the compact object nature has been solved. The lack of pulsations
may be explained by dense stellar wind ionized material smearing out the pulsed signal via free-free absorption, although the
powerful pulsar wind would allow the observer to see the pulsar without the interference of the stellar wind around inferior
conjunction. The radiation beam may also point away from us preventing us from seeing it. Regarding accretion X-ray bursts,
since the masses of the components in LS~5039 and LS~I~+61~303 are not yet properly constraint, we cannot tell whether the
compact object is a black hole or a neutron star.

Therefore, the question whether there is an accreting or a non-accreting compact object in LS~5039 and LS~I~+61~303 remains
open. Fortunately, because of this fact, their complex multiwavelength spectral and temporal behavior, and their TeV
detections, these two sources are extensively studied nowadays. This will bring for sure fruitful and surprising results in
the near future.

\subsection{The relevance of the primary star}

LS~5039, LS~I~+61~303, and Cygnus~X-1 contain a massive and hot OB type star which embeds the jet/accelerator/emitter with
dense matter and photon fields. Unlike low-mass microquasars, where the accretion/ejection phenomena could be naturally
reduced to the disk/jet system, in high-mass microquasars (SS~433 and Cygnus~X-3 are two additional instances of high-mass
microquasars, both being certainly peculiar), the strong photon field should play an important role in the radiation
processes via, e.g., stellar IC, photo-meson production and photo-disintegration, photon-photon absorption, secondary pair
radiation and electromagnetic cascading. In addition, the strong stellar wind is likely to have an impact in the radiation
processes via dynamical interactions with the jet/accelerator/emitter, providing targets for pp radiation, confining
relativistic particles, either secondary pairs or protons, absorbing part of the radio and X-ray photons produced in the
system, and determining the medium at large scales (and thereby influencing the processes that take place in the termination
regions of the microquasar jets). As shown along this review, numerous studies have been or are being carried out to clarify
the importance of the primary star in the mentioned processes. 

Beside the fact that the physics to extract from theoretical studies plus present and future observations can teach us a lot
about jets, acceleration and radiation processes, and stellar winds, it is also important to remark that the presence of the
star cannot be neglected when modeling phenomenologically the observations of high-mass microquasars. It seems very likely
that these systems cannot be understood as just accretion/ejection systems to first, or even zero, order of approximation,
but require a much more complex approach including hydrodynamical and magnetohydrodynamical simulations of jets and their
environment, a consistent treatment of acceleration and emission of particles, and the physics of OB stars and their winds.

\subsection{Prospects}

The future instruments like MAGIC~II, HESS~II, CTA, and the already on-flight GLAST will allow to make a big step in our
understanding of microquasars and gamma-ray emitting binaries. The better performance at low energies of all these
instruments will allow to study for instance the possible development of electromagnetic cascades (in case of weak magnetic
fields) or the secondary single-scattering IC component, giving information on the physical conditions of the VHE emitter
environment. Furthermore, radiation components below 100~GeV 
coming from regions invisible in TeV (e.g. due to severe absorption or maximum particle energy <~100~GeV), 
could also be investigated. The good sensitivity in the whole
energy range would allow the study of fast variability and accurate modeling, which is of primary importance to understand the
structure of the source, and the nature of the particles and processes behind the gamma-ray emission. Finally, a extension of
the operation energy range up to $\sim 100$~TeV (e.g. CTA) would bring the opportunity to study in detail the acceleration
processes with extreme efficiencies, like in the case of LS~5039.

\section{Summary}\label{summ}

Microquasars are sources in which non-thermal processes occur. Particle acceleration is taking place in these systems,
although the mechanisms involved are still unclear (e.g. non-relativistic and relativistic shocks, magnetic turbulence,
magnetic reconnection, etc.). The complexity of such a multiwavelength emitters, in which different radiative processes may
be relevant in the same energy range, and emission reprocessing via photon-photon absorption is common, yields the study in
detail of the underlying physics quite difficult. Nevertheless, the detection of TeV photons can help to constrain the
fundamental properties of the emitting region, since the relevant timescales are short. Because of this, Cherenkov astronomy
is allowing for the first time to really probe the accelerator/emitter in microquasars. Cygnus~X-1, observed by MAGIC, has
been found flaring during phases when photon-photon absorption is expected to be very severe, giving information on the
emitter location and the stellar magnetic field. LS~5039, detected by the Cherenkov telescope HESS as a periodical TeV
emitter, is at the moment the best laboratory to understand particle acceleration and radiative processes in galactic compact
sources. LS~I~61~+303, detected as well by MAGIC, shows TeV emission that varies along the orbit. Since geometric effects
would not play in this source a role for the orbital variability as important as in the case of LS~5039, intrinsic properties
of the emitter should also change. The same would apply to Cygnus~X-1.

We conclude that a leptonic scenario can explain the radiation from microquasars at very high energies, although hadronic
emission, energetically very demanding, cannot be discarded. In case of LS~5039 and Cygnus~X-1, acceleration of particles and
emission seem to take place in the borders of the system. Due to their relatively hard VHE spectra, the emitter magnetic
field in LS~5039 and LS~I~+61~303 should be low. Due to the magnetic field produced by an OB star, gamma-rays, although
likely photon-photon absorbed, can hardly trigger efficient electromagnetic cascades and secondary energy may be radiated
mainly via the synchrotron process.

TeV microquasars are showing us that they behave in a different way from their extragalactic analogs, the blazars, being very
much affected by the presence of the primary star and the the orbital motion.

\section*{Acknowledgments} 
The authors thank the anonymous referee for constructive comments and useful
suggestions. The authors are grateful to Felix A. Aharonian for fruitful
discussion and advice. The authors thank Anabella T. Araudo for helping to
improve the manuscript. V.B-R. gratefully acknowledges support from the
Alexander von Humboldt Foundation. V.B-R. acknowledges support by DGI of MEC
under grant AYA2007-68034-C03-01, as well as partial support by the European
Regional Development Fund (ERDF/FEDER).


\section{References}
\begin{thebibliography}{0}    
\bibitem{mirabel99} Mirabel, I.~F. \& Rodr\'iguez, L.~F. {\it ARA\&A} {\bf 37}, 409 (1999)
\bibitem{mcclintock06} McClintock, J.~E., Remillard, R.~A. Black Hole Binaries
in {\it Compact stellar X-ray sources}, eds. W. Lewin \& M. van der Klis 
(Cambridge University Press) (2006)
\bibitem{fender04a} Fender, R.~P., Belloni, T.~M., Gallo, E. {\it MNRAS}, {\bf 355}, 1105 (2004)
\bibitem{gallo03} Gallo E., Fender R.~P., Pooley G.~G. {\it MNRAS}, {\bf 344}, 60 (2003)
\bibitem{corbel03}
Corbel, S., Nowak, M.~A., Fender, R.~P., Tzioumis, A.~K., Markoff, S. {\it A\&A}, {\bf 400}, 1007
(2003)
\bibitem{fender03} Fender, R.~P., Gallo, E., Jonker, P.~G. {\it MNRAS} {\bf 343}, 99 (2003)
\bibitem{ribo05} Rib\'o, M. in {\it Future Directions in High Resolution Astronomy: The 10th Anniversary of the VLBA}, 
(ASPC, 2005) 340, 421 [astro-ph/0402134]
\bibitem{corbel02} Corbel, S., Fender, R. P., Tzioumis, A. K., {\it et al.}, {\it Science} {\bf 298}, 196 (2002)
\bibitem{bosch07} Bosch-Ramon, V., {\it Ap\&SS} {\bf 309}, 321 (2007)
\bibitem{fender04} Fender, R. {\it NewAR} {\bf 48}, 1399 (2004)
\bibitem{chaty03} Chaty, S., Haswell, C. A., Malzac, J. {\it MNRAS} {\bf 346}, 689 (2003)
\bibitem{levinson96} Levinson, A. \& Blandford, R.~D. {\it ApJ} {\bf 456}, L29 (1996)
\bibitem{levinson96b}Levinson, A. \& Mattox, J.~R. {\it ApJ}, {\bf 462}, 67 (1996)
\bibitem{paredes00} Paredes, J.M., Mart\'{\i}, J., Rib\'o, M.,  Massi, M., {\it Science} {\bf 288}, 2340 (2000)
\bibitem{albert07} Albert, J. et al. {\it ApJ} {\bf 665}, L51 (2007)
\bibitem{aharonian05} Aharonian, et al. {\it Science} {\bf 309}, 746 (2005)
\bibitem{aharonian06} Aharonian, F.~A. et al. {\it A\&A} {\bf 460}, 743 (2006)
\bibitem{casares05} Casares, J., Rib\'o, M., Ribas, I., et al., {\it MNRAS} {\bf 364}, 899 (2005)
\bibitem{albert06} Albert, J. et al. {\it Science} {\bf 312}, 1771 (2006)
\bibitem{mirabel92} Mirabel I. F., Rodr\'iguez L. F., Cordier B., Paul J., Lebrun F., {\it Nature} {\bf 358}, 215 (1992)
\bibitem{mirabel94} Mirabel, I. F. \& Rodriguez, L. F. {\it Nature} {\bf 371}, 46 (1994)
\bibitem{merloni03} Merloni, A., Heinz, S., di Matteo, T. {\it MNRAS} {\bf 345}, 1057 (2003)
\bibitem{falcke04} Falcke, H., Körding, E., Markoff, S., {\it A\&A} {\bf 414}, 895 (2004)
\bibitem{koerding06} K\"ording, E., Falcke, H., Corbel, S. {\it A\&A} {\bf 456}, 439 (2006)
\bibitem{markoff05} Markoff, S., Nowak, M.~A., Wilms, J., {\it ApJ} {\bf 635}, 1203 (2005)
\bibitem{maccarone05} Maccarone, T.~J., {\it MNRAS} {\bf 360}, L68 (2005)
\bibitem{bogovalov05} Bogovalov, S.~V. \& Kelner, S.~R. {\it Astron. Rep.} {\bf 49}, 57 (2005)
\bibitem{levinson06} Levinson, A. {\it Int. J. Mod. Phys.~A} {\bf 21}, 30 (2006)
\bibitem{araudo07} Araudo, A.~T., Romero, G.~E., Bosch-Ramon, V., Paredes, J.~M. {\it A\&A} {\bf 476}, 1289 (2007)
\bibitem{blandford76} Blandford, R. D. {\it MNRAS} {\bf 176} 465 (1976)
\bibitem{blandford77} Blandford, R. D. \& Znajek, R. L. {\it MNRAS} {\bf 179}, 433 (1977)
\bibitem{blandford82} Blandford, R. D. \& Payne, D. G. {\it MNRAS} {\bf 199}, 883 (1982) 
\bibitem{meier96} Meier, D. {\it ApJ}, {\bf 459}, 185 (1996)
\bibitem{koide02} Koide, S., Shibata, K., Kudoh, T., Meier, D.~L. {\it Science} {\bf 295}, 1688 (2002)
\bibitem{chatto02} Chattopadhyay, I. \& Chakrabarti, S.~K. {\it MNRAS} {\bf 333}, 454 (2002)
\bibitem{meier03} Meier, D. {\it NewA Rev.}, {\bf 47}, 667 (2003)
\bibitem{hujeirat04} Hujeirat, A. {\it A\&A} {\bf 416}, 423--435 (2004)
\bibitem{meier05} Meier, D.~L. {\it Ap\&SS} {\bf 300}, 55 (2005)
\bibitem{deVilliers05} De Villiers, J.-P., Hawley, J.~F., Krolik, J.~H., Hirose, S. {\it ApJ} {\bf 620}, 878 (2005)
\bibitem{ferreira06} Ferreira, J., Petrucci, P.~O., Henri, G., Saugé, L., Pelletier, G. {\it A\&A} {\bf 447}, 813 (2006)
\bibitem{mcKinney06} McKinney, J.~C. {\it MNRAS}, {\bf 368}, 1561 (2006)
\bibitem{hawley06}  Hawley, J.~F. \& Krolik, J.~H. {\it ApJ} {\bf 641}, 103 (2006)
\bibitem{komissarov07} Komissarov, S.~S., Barkov, M.~V., Vlahakis, N., K\"onigl, A. {\it MNRAS} {\bf 380}, 51 (2007)
\bibitem{barkov08} Barkov, M.~V. \& Komissarov, Serguei S. {\it MNRAS} {\bf 385}, 28 (2008)
\bibitem{junor99} Junor, W., Biretta, J.~A., Livio, M. {\it Nature} {\bf 401}, 891 (1999)
\bibitem{horiuchi06} Horiuchi, S., Meier, D.~L., Preston, R.~A., Tingay, S.~J. {\it PASJ} {\bf 58}, 221 (2006)
\bibitem{namiki03} Namiki, M., Kawai, N., Kotani, T., Makishima, K., {\it PASJ} {\bf 55}, 281 (2003)
\bibitem{hardee03} Hardee, P.~E., Hughes, P.~A. {\it ApJ} {\bf 583}, 116 (2003)
\bibitem{tsinganos04} Tsinganos, K., Vlahakis, N., Bogovalov, S.~V., Sauty, C., Trussoni, E. {\it Ap\&SS} {\bf 293}, 55 (2004)
\bibitem{perucho07} Perucho, M. \& Bosch-Ramon, V. {\it A\&A} {\bf 482}, 917 (2008)  
\bibitem{faranoff74} Fanaroff, B.~L. \& Riley, J.~M. {\it MNRAS} {\bf 167}, 31 (1974)
\bibitem{pm07} Perucho, M. \& Mart\'{\i}, J.~M. {\it MNRAS} {\bf 382}, 526 (2007)
\bibitem{kaiser97} Kaiser, C.~R. \& Alexander, P. {\it MNRAS} {\bf 286}, 215 (1997)
\bibitem{sch02} Scheck, L., Aloy, M.A., Mart\'{\i}, J.M$^{\underline{\rm a}}$, G\'omez, J.L., M\"uller, E. {\it MNRAS} {\bf 331}, 615 (2002)
\bibitem{zealey80}  Zealey, W.~J., Dopita, M.~A., Malin, D.~F. {\it MNRAS} {\bf 192}, 731 (1980)
\bibitem{velazquez00} Vel\'azquez, P. F. \& Raga, A. C. {\it A\&A}, {\bf 362}, 780 (2000) 
\bibitem{sikora05} Sikora, M., Begelman, M.~C., M., Greg M., Lasota, J.~P., {\it ApJ} {\bf 625}, 72 (2005)
\bibitem{zenitani01} Zenitani, S., Hoshino, M., {\it ApJ} {\bf 562}, 63 (2001)
\bibitem{derishev03} Derishev, E.~V., Aharonian, F.~A., Kocharovsky, V.~V., Kocharovsky, Vl. V., 
{\it PhRvD} {\bf 68}, 3003 (2003)
\bibitem{stern06} Stern, B., Poutanen, J. {\it MNRAS} {\bf 372}, 1217 (2006)
\bibitem{gierlinski03} Gierli\'nski, M., Done, C., {\it MNRAS} {\bf 342}, 1083 (2003)
\bibitem{neronov07} Neronov, A. \& Aharonian, F.~A. {\it ApJ} {\bf 671}, 85 (2007)
\bibitem{rieger07} Rieger, F. \& Aharonian, F.~A. {\it A\&A} {\bf 479}, L5 (2008)
\bibitem{drury83} Drury, L.~O., {\it Rep. Prog. Phys.} {\bf 46}, 973 (1983)
\bibitem{fermi49} Fermi, E., {\it Phys. Rev.} {\bf 75}, 1169 (1949)
\bibitem{rieger04} Rieger, F.~M.; Duffy, P., {\it ApJ} {\bf 617}, 155 (2004)
\bibitem{rieger06} Rieger, F.~M., Bosch-Ramon, V., Duffy, P. {\it Ap\&SS} {\bf 309}, 119 (2007)
\bibitem{heinz02} Heinz, S., Sunyaev, R., {\it A\&A} {\bf 390}, 751 (2002)
\bibitem{markoff01} Markoff S., Falcke H., Fender R., {\it A\&A} {\bf 372}, 25 (2001)
\bibitem{bosch06b} Bosch-Ramon, V., Romero, G~E., Paredes, J.~M. {\it A\&A} {\bf 447}, 263 (2006)
\bibitem{bosch04} Bosch-Ramon, V., Paredes, J. M., {\it A\&A} {\bf 417}, 1075 (2004)
\bibitem{romero02} Romero, G.~E., Kaufman Bernad\'o, M.~M., Mirabel, I.~F., {\it A\&A} {\bf 393}, 61 (2002)
\bibitem{georganopoulos02} Georganopoulos, M., Aharonian, F.~A., Kirk, J.~G., {\it A\&A} {\bf 388}, 25 (2002)
\bibitem{levinson01} Levinson, A., Waxman, E. {\it PhRvL} {\bf 87}, 171101 (2001)
\bibitem{aharonian06b} Aharonian, F.~A., Anchordoqui, L.~A., Khangulyan, D., Montaruli, T. {\it J.Phys.Conf.Ser.} {\bf 39} 408 (2006)
\bibitem{romero08} Romero, G.~E. \& Vila G.~S. {\it A\&A} {\bf 485}, 623 (2008) 
\bibitem{yuan05} Yuan, F.; Cui, W., Narayan, R., {\it ApJ} {\bf 620}, 905 (2005)
\bibitem{paredes06} Paredes, J.~M., Bosch-Ramon, V., Romero, G.~E., {\it A\&A} {\bf 451}, 259 (2006)
\bibitem{atoyan99} Atoyan, A.~M., Aharonian, F.~A., {\it MNRAS} {\bf 302}, 253 (1999)
\bibitem{dermer06} Dermer, C., B\"ottcher, M., {\it ApJ} {\bf 643}, 1081 (2006)
\bibitem{paredes02} Paredes, J. M., Rib\'o, M., Mart\'i, J., {\it A\&A} {\bf 393}, 99 (2002)
\bibitem{kaufman02} Kaufman Bernad\'o M.M., Romero G.E., Mirabel I.F., {\it A\&A} {\bf 385}, 10 (2002)
\bibitem{khangulyan07} Khangulyan, D., Aharonian, F., Bosch-Ramon, V., {\it MNRAS}, {\bf 383}, 467 (2008)
\bibitem{romero03} Romero, G. E., Torres, D. F., Kaufman Bernad\'o, M. M., Mirabel, I. F., {\it A\&A} {\bf 410}, 1 (2003)
\bibitem{romero05} Romero, G.~E., Orellana, M., {\it A\&A} {\bf 439}, 237 (2005)
\bibitem{aharonian06c} Aharonian, F. Invited talk at {\it The International Conference on Neutrino Physics and Astrophysics} (2006)
[astro-ph/0702680]
\bibitem{bednarek05} Bednarek, W. {\it ApJ} {\bf 631}, 466 (2005)
\bibitem{vanderlaan66} van der Laan, H., {\it Nature} {\bf 211}, 1131 (1966)
\bibitem{aharonian98} Aharonian, F. A.; Atoyan, A. M. {\it NewAR} {\bf 42}, 579 (1998)
\bibitem{bordas08} Bordas, P., Bosch-Ramon, V., Paredes, J. M. {\it Int.~J.~Mod Phys.~D} {\bf 17}, 1895 (2008)
\bibitem{bosch05} Bosch-Ramon, V., Aharonian, F. A., Paredes, J. M., {\it A\&A} {\bf 432}, 609 (2005c)
\bibitem{akharonian85} Akharonian, F. A., Vardanian, V. V., {\it Ap\&SS} {\bf 115}, 31 (1985)
\bibitem{donati02} Donati, J.~F., Babel, J., Harries, T.~J. et al. {\it MNRAS} {\bf 33}, 55 (2002)
\bibitem{protheroe87} Protheroe, R.~J. \& Stanev, T. , {\it ApJ} {\bf 322}, 838 (1987)
\bibitem{moskalenko94} Moskalenko I.V., Karakula S., {\it ApJS} {\bf 92}, 567 (1994)
\bibitem{bednarek97} Bednarek, W. {\it A\&A}, {\bf 322}, 523 (1997)
\bibitem{boettcher05} Böttcher, M., Dermer, C.~D. {\it ApJ}, {\bf 634}, L81 (2005)
\bibitem{bottcher05} B\"ottcher, M., Dermer, C., {\it ApJ} {\bf 634}, 81 (2005)
\bibitem{dubus06} Dubus, G., {\it A\&A} {\bf 451}, 9 (2006)
\bibitem{reynoso08} 
Reynoso, M.~M., Christiansen, H.~R., Romero, G. E. {\it Astrop. Phys.} {\bf 28}, 565 (2008)
\bibitem{bednarek06} Bednarek, W., {\it MNRAS} {\bf 368}, 579 (2006)
\bibitem{orellana07} Orellana, M., Bordas, P., Bosch-Ramon, V., Romero, G.~E., Paredes, J.~M. {\it A\&A} {\bf 476}, 9 (2007) 
\bibitem{bosch08} Bosch-Ramon, V., Khangulyan, D., Aharonian, F.~A. {\it A\&A} {\bf 482}, 397 (2008)
\bibitem{hillas84} Hillas, A.~M. {\it ARA\&A} {\bf 22} 425 (1984)
\bibitem{protheroe99} Protheroe, R.~J. Acceleration and Interaction of Ultra High Energy Cosmic Rays in 
{\it Topics in cosmic-ray astrophysics}, eds. M. A. DuVernois (Nova Science Publishing) (1999) [astro-ph/9812055]
\bibitem{kelner06} Kelner, S.~R., Aharonian, F.~A., Bugayov, V.~V. {\it Phys. Rev. D} {\bf 74}, 4018 (2006)
\bibitem{kelner08} Kelner, S. R. \& Aharonian, F. A. {\it Phys. Rev. D} {\bf 78}, 4013 (2008)  
\bibitem{aharonian98b} Aharonian F.~A. \& Heinzelmann, G. {\it Nucl. Phys. Proc. Suppl.} {\bf 60}, 193 (1998)
\bibitem{rovero02} Rovero A.~C. et al., {\it BAAA} {\bf 45}, 66 (2002)
\bibitem{khangulyan05} Khangulyan, D. \& Aharonian, F.~A. in {\it High Energy Gamma-Ray Astronomy}, 
(AIP Conference Proceedings 2005), 745, 359 [astro-ph/0503499]
\bibitem{stirling01} Stirling, A. M., Spencer, R. E., de la Force, C. J. {\it MNRAS}, {\bf 327}, 1273 (2001)
\bibitem{ziolkowski05} Ziolkowski, J. {\it MNRAS}, {\bf 358}, 851 (2005)
\bibitem{gies86} Gies, D.~R., \& Bolton, C. T. {\it ApJ}, {\bf 304}, 371 (1986)
\bibitem{bednarek07} Bednarek, W. \& Giovannelli, F. {\it A\&A} {\bf 464}, 437 (2007)
\bibitem{sunyaev79} Sunyaev, R. A., \& Truemper, J. {\it Nature} {\bf 279}, 506 (1979)
\bibitem{mcconnell02} McConnell, M. L., Zdziarski, A. A., Bennett, K. {\it ApJ} {\bf 572}, 984 (2002)
\bibitem{bosch08b} Bosch-Ramon, V., Khangulyan, D., Aharonian, F. A. {\it A\&A} {\bf 489}, L21  (2008) 
\bibitem{motch97} Motch, C., Haberl, F., Dennerl, K., Pakull, M., Janot-Pacheco, E., {\it A\&A} {\bf 323}, 853 (1997)
\bibitem{bosch05} Bosch-Ramon, V., Paredes, J. M., Rib\'o, M. et al. {\it ApJ} {\bf 628}, 388 (2005)
\bibitem{martocchia05} Martocchia, A., Motch, C., Negueruela, I., {\it A\&A} {\bf 430}, 245 (2005)
\bibitem{ribo99} Rib\'o, M., Reig, P., Mart\'i, J., Paredes, J. M., {\it A\&A}, {\bf 347}, 518 (1999)
\bibitem{ribo08} Rib\'o, M., Paredes, J. M., Moldon, J., Mart\'i, J., Massi, M. {\it A\&A} {\bf 481}, 17 (2008)
\bibitem{ribo02} Rib\'o, M. Ph. D. Thesis, Universitat de Barcelona (2002)
\bibitem{marti98} Mart\'i, J., Paredes, J. M., Rib\'o, M., {\it A\&A} {\bf 338}, 71 (1998)
\bibitem{horns06} Horns, D., for the HESS collaboration, talk presented in 2nd Workshop On TeV Particle Astrophysics (2006)
\bibitem{bosch07b} Bosch-Ramon, V., Motch, C., Rib\'o, M., et al. {\it A\&A} {\bf 473}, 545 (2007)
\bibitem{denaurois06} de Naurois, M. for the HESS collaboration, talk presented in The keV to TeV connection (2006)
\bibitem{dubus07} Dubus, G., Cerutti, B., Henri, G. {\it A\&A} {\bf 477}, 691 (2008)
\bibitem{sierpowska07} Sierpowska-Bartosik, A., Torres, D.~F. {\it ApJL}, {\bf 671}, 145 (2007)
\bibitem{sierpowska08} Sierpowska-Bartosik, A., Torres, D.~F. {\it ApJL}, {\bf 674}, 89 (2008)
\bibitem{hutchings81} Hutchings, J. B. \& Crampton, D. {\it PASP} {\bf 93}, 486 (1981)
\bibitem{gregory02} Gregory, P. C. {\it ApJ}, {\bf 575}, 427 (2002)
\bibitem{casares05b} Casares, J., Ribas, I., Paredes, J. M., Mart\'i, J., Allende Prieto, C. {\it MNRAS} {\bf 360}, 1105 (2005)
\bibitem{massi01} Massi, M., Rib\'o, M., Paredes, J. M., Peracaula, M., Estalella, R. {\it A\&A} {\bf 376}, 217 (2001)
\bibitem{massi04} Massi, M., Rib\'o, M., Paredes, J. M., et al. {\it A\&A} {\bf 414}, L1 (2004)
\bibitem{sidoli06} Sidoli, L., Pellizzoni, A., Vercellone, S., et al. {\it A\&A} {\bf 459}, 901 (2006)
\bibitem{paredes07} Paredes, J. M., Rib\'o, M., Bosch-Ramon, V., et al. {\it ApJ} {\bf 664}, 39 (2007)
\bibitem{dhawan06} Dhawan, V., Mioduszewski, A., Rupen, M. The VI Microquasar Workshop: Microquasars and Beyond (Proceedings of Science), 52, 1 (2006)
\bibitem{dubus06b} Dubus, G. {\it A\&A} {\bf 456}, 801 (2006)
\bibitem{chernyakova06} Chernyakova, M., Neronov, A., Walter, R. {\it MNRAS} {\bf 372}, 1585 (2006)
\bibitem{maraschi81} Maraschi, L. \& Treves, A. {\it MNRAS} {\bf 194}, 1 (1981)
\bibitem{romero07} Romero, G.~E., Okazaki, A.~T., Orellana, M., Owocki, S.~P. {\it A\&A} {\bf 474}, 15 (2007)
\bibitem{bogovalov08} Bogovalov, S.~V., Khangulyan, D., Koldoba, A.~V., Ustyugova, G.~V., Aharonian, F.~A. {\it MNRAS} 
{\bf 387}, 63 (2008) 
\bibitem{kniffen97} Kniffen, D.~A., Alberts, W.~C.~K., Bertsch, D.~L. et al. {\it ApJ} {\bf 486} 126 (1997)
\bibitem{romero05b} Romero, G. E., Christiansen, H. R., Orellana, M. {\it ApJ} {\bf 632}, 1093 (2005)
\bibitem{bosch06c} Bosch-Ramon, V., Paredes, J.~M., Romero, G. E., Rib\'o, M. {\it A\&A} {\bf 459}, L25 (2006)
\bibitem{leahy04} Leahy, D. A., {\it A\&A}, {\bf 413}, L1019 (2004)
\bibitem{christiansen06} Christiansen, H.~R., Orellana, M., Romero, G.~E. {\it Phys. Rev. D.} {\bf 73}, 3012 (2006)
\bibitem{torres07} Torres, D.~F. \& Halzen F. {\it Astropart.Phys.} {\bf 27}, 500 (2007)
\bibitem{grundstrom07} Grundstrom, E.~D., Caballero-Nieves, S.~M., Gies, D.~R. et al. {\it ApJ} {\bf 656}, 437 (2007)
\end{thebibliography}
\end{document}